\documentclass[12pt,letterpaper]{article}
\usepackage{epsfig,rotating,setspace,latexsym,amsmath,epsf,amssymb,amsfonts,bm,theorem,cite,caption,subcaption,enumerate,longtable,accents}
\usepackage{algorithm,algorithmic,graphicx,epsf,authblk,epstopdf,url,color,multirow}

\newtheorem{theorem}{Theorem}

\newtheorem{remark}{Remark}
\newtheorem{lemma}{Lemma}
\newenvironment{Proof}[1]{\medskip\par\noindent{\bf Proof:\,}\,#1}{{\mbox{\,$\blacksquare$}\par}}

\newcommand{\cp}{{\mathcal{P}}}

\allowdisplaybreaks

\begin{document}

\title{Multi-Party Private Set Intersection: An Information-Theoretic Approach\thanks{This work was supported by NSF Grants CCF 17-13977 and ECCS 18-07348.}}

\author[1]{Zhusheng Wang}
\author[2]{Karim Banawan}
\author[1]{Sennur Ulukus}

\affil[1]{\normalsize Department of Electrical and Computer Engineering, University of Maryland}
\affil[2]{\normalsize Electrical Engineering Department, Faculty of Engineering, Alexandria University}

\maketitle

\vspace*{-0.8cm}

\begin{abstract}
We investigate the problem of multi-party private set intersection (MP-PSI). In MP-PSI, there are $M$ parties, each storing a data set $\cp_i$ over $N_i$ replicated and non-colluding databases, and we want to calculate the intersection of the data sets $\cap_{i=1}^M \cp_i$ without leaking any information beyond the set intersection to any of the parties. We consider a specific communication protocol where one of the parties, called the leader party, initiates the MP-PSI protocol by sending queries to the remaining parties which are called client parties. The client parties are not allowed to communicate with each other. We propose an information-theoretic scheme that privately calculates the intersection $\cap_{i=1}^M \cp_i$ with a download cost of $D = \min_{t \in \{1, \cdots, M\}}  \sum_{i \in \{1, \cdots M\}\setminus {t}} \left\lceil \frac{|\cp_t|N_i}{N_i-1}\right\rceil$. Similar to the 2-party PSI problem, our scheme builds on the connection between the PSI problem and the multi-message symmetric private information retrieval (MM-SPIR) problem. Our scheme is a non-trivial generalization of the 2-party PSI scheme as it needs an intricate design of the shared common randomness. Interestingly, in terms of the download cost, our scheme does not incur any penalty due to the more stringent privacy constraints in the MP-PSI problem compared to the 2-party PSI problem.
\end{abstract}

\section{Introduction}
The two-party private set intersection (PSI) problem refers to a classical privacy problem, which is introduced in \cite{PSI_first}. In its classical setting, two parties, each possessing a data set, need to calculate common elements that lie in both data sets. This calculation is performed in such a way that neither party reveals anything to the counterparty except for the elements in the intersection. Ubiquitous schemes have been investigated to tackle the PSI problem using cryptographic techniques; see for example \cite{PSI_computational, PSI_efficient, PSI_survey}. Many practical applications are tied to PSI. To see this, consider the following scenario: Suppose that the national security agency (NSA) and the customs and border protection (CBP) need to check whether a specific group of suspected criminals has entered the country. The NSA has a list of suspected criminals, while the CBP has a complete list of individuals who entered the country. Both agencies want to find the intersection between these lists. However, the NSA does not want to share its complete list of suspects, and the CBP cannot reveal the entire catalog of records either. This is a natural application for the 2-party PSI problem. Reference \cite{PSIjournal} formulates the 2-party PSI problem from an information-theoretic perspective. Interestingly, \cite{PSIjournal} explores an intriguing connection between the PSI problem and the private information retrieval (PIR) problem \cite{PIR_ORI}. Specifically, \cite{PSIjournal} investigates the PSI determination using the multi-message symmetric PIR (MM-SPIR) procedure. Surprisingly, under some technical conditions, MM-SPIR proves to be the most-efficient PSI protocol under absolute privacy guarantees. The efficiency is measured by the total download cost, which is the number of bits needed to be downloaded to calculate the set intersection at one of the parties. The optimality proof builds on the rich literature of characterizing the fundamental limits of PIR and related problems, starting with the seminal work of Sun-Jafar \cite{PIR}. Further fundamental limits of many variations of the PIR problem have been investigated; see \cite{JafarColluding, arbitraryCollusion, RobustPIR_Razane, Staircase_PIR, SPIR, codedsymmetric, wang2017linear, SPIR_Mismatched, ChaoTian_leakage, KarimCoded, codedcolluded, codedcolludingZhang, Kumar_PIRarbCoded, codedcolludingJafar, MM-PIR, MPIRcodedcolludingZhang, BPIRjournal, CodeColludeByzantinePIR, tandon2017capacity, KimCache, wei2017fundamental, wei2017fundamental_partial, PIR_cache_edge, kadhe2017private, chen2017capacity, wei2017capacity, MMPIR_PSI, SSMMPIR_SI1, SSMMPIR_SI2, LiConverse, StorageConstrainedPIR_Wei, PrivateComputation, mirmohseni2017private, PrivateSearch, abdul2017private, StorageConstrainedPIR, efficient_storage_ITW2019, Chao_storage_cost, PIR_decentralized, heteroPIR, TamoISIT, Karim_nonreplicated, PIR_WTC_II, SecurePIR, securePIRcapacity, securestoragePIR,  XSTPIR, arbmsgPIR, ChaoTian_coded_minsize, MultiroundPIR, KarimAsymmetricPIR, noisyPIR, PIR_lifting, PIR_networks} for example. 

The MM-SPIR framework to solve the PSI problem in \cite{PSIjournal}, however, works only for 2-party PSI. This is because the original PIR problem (and the SPIR problem) involves two parties, the user and the server(s). Unlike PIR, the PSI problem may involve more than two parties. Returning to the example involving the NSA and CBP above, suppose now that the NSA needs to narrow down the search to check whether the suspects have entered the country via a specific airline. The airline company has a list of all passengers that took its flights all over the world. The company needs to protect the privacy of its passengers as well. The problem of finding the set of suspects who entered the country via this specific airline becomes a 3-party PSI. Unfortunately, the NSA cannot just apply a 2-party PSI scheme with the airline company and the CBP, as the NSA will learn extra information than the intersection of the three lists, for example, the NSA will learn about some of its suspects who boarded a flight with this airline company but never landed in this country. Another example of 3-party PSI is related to ad clicks. Consider a company which sells a certain product (e.g., shoes), a company which makes ads and posts them at various web-hosts, and another company which is a web-host that hosts ads. All of these parties have their individual lists of clicks that they wish to keep private, but may want to compute the intersection, i.e., actual customers who bought the product from the company after seeing an ad produced by the ad company hosted at the particular web-host company, to determine the effectiveness of the ad company and the web-host company. Note again that pairwise intersections leak additional information beyond the three-way intersection. These examples motivate the multi-party PSI (MP-PSI) problem. They also illustrate that the MP-PSI is a non-trivial extension of the 2-party PSI as it cannot be implemented via multiple 2-party PSI. To make this point even stronger, consider three parties with sets $\mathcal{P}_1=\{1, 2, 3\}$, $\mathcal{P}_2=\{1, 2\}$ and $\mathcal{P}_3=\{1, 3\}$. The intersection of these three sets is $\mathcal{P}=\mathcal{P}_1 \cap \mathcal{P}_2 \cap \mathcal{P}_3=\{1\}$. When any one these parties is chosen as the leader party and applies a 3-party PSI protocol, the leader party should learn only this three-way intersection. However, if the leader party applies a 2-party PSI with the two client parties, it will learn information more than the three-way intersection. For instance, if the leader party is the first party, and if it applies a 2-party PSI with the second and third parties, it will learn $\mathcal{P}_1 \cap \mathcal{P}_2=\{1, 2\}$ and $\mathcal{P}_1 \cap \mathcal{P}_3=\{1, 3\}$. Even though the leader party can obtain the three-way intersection by taking the intersection of these two two-way intersections, i.e, $\{1, 2\} \cap \{1, 3\} = \{1\}$, this sequential use of 2-party PSI for the 3-party PSI problem leaks further information to the leader party. For instance, the leader party learns that the second party has $\{2\}$ and the third party has $\{3\}$ further than the overall intersection $\{1\}$. Thus, 3-party PSI cannot be implemented by two 2-party PSI. In the computational privacy literature, the first MP-PSI achievable scheme was proposed by Freedman et al. \cite{FNP04}. Though considerable progress has been made in the construction of various 2-party PSI schemes, only few works exist for MP-PSI schemes \cite{HV17}. 

\begin{figure}[t]
	\centering
	\epsfig{file=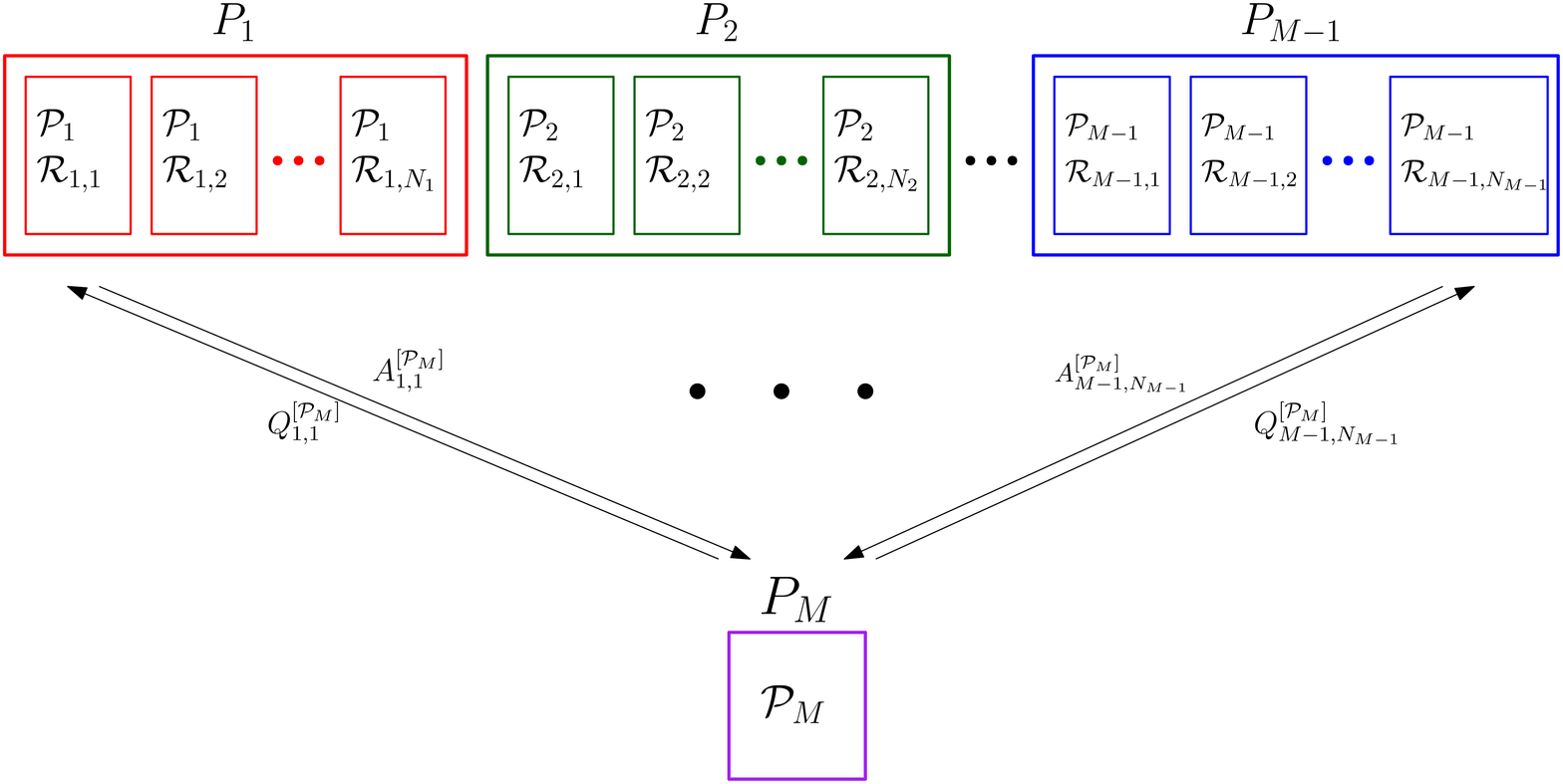,width=1\textwidth}
	\caption{Multi-party private set intersection (MP-PSI) system model.}
	\label{MP_PSI_Setting}
\end{figure}

In this paper, we investigate the MP-PSI problem from an information-theoretic perspective. In MP-PSI, there are $M$ independent parties. The $i$th party is denoted by $P_i$, for $i=1, \cdots, M$. Each party possesses a data set $\cp_i$, where $i \in \{1, \cdots, M\}$. The elements of all data sets are picked from a finite set $\mathbb{S}_K$ with cardinality $|\mathbb{S}_K|=K$ for sufficiently large $K$\footnote{Without loss of generality, one can assume that $\mathbb{S}_K=\{1, \cdots, K$\}.}. The data set $\cp_i$ is stored in $N_i$ replicated and non-colluding databases. We aim at privately determining the intersection of all the $M$ data sets, i.e., we aim at calculating  $\mathcal{P} = \cap_{i=1}^M \cp_i$ in such a way that no party can learn any information beyond the intersection $\cp$. Inspired by the classical achievable scheme in \cite{FNP04, KS05}, we focus on a specific communication strategy between the parties in this work; see Fig.~\ref{MP_PSI_Setting}. In particular, we assume that the parties agree on choosing one of them as a \emph{leader} party, while the remaining parties act as \emph{client} parties. Without loss of generality, we pick $P_M$ as a leader party, and then the remaining parties $P_1,\cdots, P_{M-1}$ are all client parties. The leader party $P_M$ initiates the MP-PSI determination protocol by generating and submitting queries to the client parties. At the clients' side and before MP-PSI, the clients are allowed to generate and share common randomness (common randomness residing in the $j$th database of party $P_i$ is shown by $\mathcal{R}_{i,j}$ in Fig.~\ref{MP_PSI_Setting}). This is motivated by the results of \cite{PSIjournal, SPIR, SPIR_ORI}, which assert that using common randomness is strictly necessary to enable symmetrically private communication. Furthermore, we assume that the leader party $P_M$ can communicate with each client party in only one round, and communication between any two client parties is not allowed during the protocol. The client parties respond truthfully to the leader's queries without leaking information about the elements outside $\cp$ with the aid of the assigned common randomness.       

In this paper, we first formulate the MP-PSI problem from an information-theoretic perspective. We show that MP-PSI can also be recast as a MM-SPIR problem, which extends the formulation of the 2-party PSI problem \cite{PSIjournal}. This can be done by mapping the data sets at each party into an \emph{incidence vector}\footnote{As investigated in \cite{PSIjournal}, in MM-SPIR problem, a user needs to retrieve $P$ messages from $N$ replicated servers containing $K$ messages. The PSI problem can be recast as MM-SPIR by considering that the messages correspond to \emph{incidences} of elements in its data set with respect to the finite set of all elements. Specifically, \cite{PSIjournal} transforms each data set into a library of $K$ binary messages of one-bit length. Finally, in \cite{PSIjournal}, party $P_1$ (or $P_2$) performs MM-SPIR of the messages corresponding to its data set $\cp_1$ (or $\cp_2$) within the databases of the other entity.} to facilitate the MM-SPIR of the elements that belong to $\cp_M$ (the leader's data set). Next, we propose a novel achievable scheme for MP-PSI determination. The structure of the queries that the leader party submits in our scheme is the same as the SPIR queries in \cite{SPIR} (which can be traced back to the original work of Chor et al. in \cite{PIR_ORI}). Despite the similarity of the queries, the answering strings in MP-PSI are fundamentally different. This is due to the fact that the leader party cannot perform $M-1$ pair-wise PSI operation to calculate $\cp=\cap_{i=1}^{M} \cp_i$ without leaking extra information about the individual intersections $\cp_M \cap \cp_i$,  $i=1, \cdots, M-1$, as discussed above. Note that, in general the joint intersection set $\cp \neq \cup_{i=1}^{M-1} \left(\cp_M \cap \cp_i\right)$. To alleviate this problem, we design an intricate protocol of generating and sharing the common randomness among the databases of the parties. By properly incorporating the common randomness to the answer strings, we prevent the leader party from learning about the elements that lie in $\cup_{i=1}^{M-1} \left(\cp_M \cap \cp_i\right)$ but not in $\cap_{i=1}^{M} \cp_i$. This constraint is referred to as the clients' privacy constraint. By correlating some of the components of the common randomness in a specific way, we show that the leader party can reliably identify the elements in $\cp$, but nothing beyond it. The download cost of our scheme is $\min_{t \in \{1, \cdots, M\}}  \sum_{i \in \{1, \cdots M\}\setminus {t}} \left\lceil \frac{|\cp_t|N_i}{N_i-1}\right\rceil$. This means that although in MP-PSI, the clients' privacy constraint is more stringent than that in the PSI problem, we incur no penalty for it. In addition, our achievable download cost scales linearly with the cardinality of the leader set, which outperforms the best-known MP-PSI scheme, which scales with the sum of the cardinalities of the data sets \cite{HV17}. Furthermore, our scheme has an advantage of simpler implementation in addition to providing absolute (information-theoretic) privacy guarantees compared to the computationally private techniques in the literature.

\section{Problem Formulation}\label{MP-PSI-Formulation}
Consider a setting where there are $M$ independent parties, denoted by $P_i,\: i=1,2,\cdots, M$. The $i$th party possesses a data set $\mathcal{P}_i$ for $i \in [1:M]$. The data set $\cp_i$ is stored within $N_i$ replicated and non-colluding databases. Given that $K$ is large enough, the elements in each data set $\cp_i$ are picked from a finite set $\mathbb{S}_K$ of cardinality $K$ with an arbitrary statistical distribution\footnote{The presented achievability scheme works for any data set generation model. The specific data set generation model in the 2-party PSI problem in \cite{PSIjournal} was  introduced only for settling the converse.}. We assume that the cardinality of data set $|\mathcal{P}_i|$ is public knowledge.

Motivated by the relation between 2-party PSI and MM-SPIR in \cite{PSIjournal}, the $i$th party maps its data set $\cp_i$ into a searchable list to facilitate PIR. To that end, the party $P_i$ constructs an incidence vector $X_i$, which is a binary vector of size $K$ associated with the data set $\mathcal{P}_i$ for all $i \in [1:M]$, such that
\begin{align}
X_{i,j}=\begin{cases}
1, \quad j \in \cp_i \\
0, \quad j \notin \cp_i
\end{cases}
\end{align}
where $X_{i,j}$ is the $j$th element of $X_i$ for all $j \in \mathbb{S}_K$. Note that $X_i$ is a sufficient statistic for $\mathcal{P}_i$ for a given $K$. Hence, the MP-PSI determination is performed over $X_i$ instead of $\mathcal{P}_i$. 

We consider a specific communication protocol in this work. The parties agree on a \emph{leader} party, which sends queries to the remaining parties and eventually calculates the desired intersection $\cap_{i=1}^M \cp_i$. The remaining parties are called \emph{client} parties. Without loss of generality, assume that the leader party is $P_M$. The leader party $P_M$ sends the query $Q_{i,j}^{[\mathcal{P}_M]}$ to the $j$th database in the client party $P_i$ for all $i \in [1:M-1]$ and $j \in [1:N_i]$. Since $P_M$ has no information about data set $\mathcal{P}_i$ before the communication, the generated queries $Q_{i,j}^{[\mathcal{P}_M]}$ are independent from $\mathcal{P}_i$. Hence,
\begin{align} \label{Independence}
I(Q_{i,j}^{[\mathcal{P}_M]};\mathcal{P}_i)=0, \quad \forall i  \in [1:M-1], \forall j \in [1:N_i]
\end{align}  

The $j$th database associated with the client party $P_i$ responds truthfully with an answer  $A_{i,j}^{[\mathcal{P}_M]}$ for all $i \in [1:M-1]$, and $j \in [1:N_i]$. The answer is a deterministic function of the query $Q_{i,j}^{[\mathcal{P}_M]}$, the data set $\mathcal{P}_i$, and some common randomness $\mathcal{R}_{i,j}$ that is available to the $j$th database of $P_i$. Thus,
\begin{align} \label{Determined Anwser}
H(A_{i,j}^{[\mathcal{P}_M]}|Q_{i,j}^{[\mathcal{P}_M]},\mathcal{P}_i,\mathcal{R}_{i,j})=0, \quad \forall i  \in [1:M-1], \forall j \in [1:N_i]
\end{align} 

Let us denote all the queries generated by $P_M$ as $Q_{1:M-1,1:N_i}^{[\mathcal{P}_M]}$ and all the answers collected by $P_M$ as $A_{1:M-1,1:N_i}^{[\mathcal{P}_M]}$, i.e.,
\begin{align}
Q_{1:M-1,1:N_i}^{[\mathcal{P}_M]} = \left \{Q_{i,j}^{[\mathcal{P}_M]}: i \in [1:M-1], j \in [1:N_i] \right \} \\
A_{1:M-1,1:N_i}^{[\mathcal{P}_M]} = \left \{A_{i,j}^{[\mathcal{P}_M]}: i \in [1:M-1], j \in [1:N_i] \right \}
\end{align}

Three formal requirements are needed to be satisfied for the MP-PSI problem:

First, the leader party $P_M$ should be able to reliably determine the intersection $\mathcal{P}=\cap_{i=1}^M \cp_i$ based on  $Q_{1:M-1,1:N_i}^{[\mathcal{P}_M]}$, $A_{1:M-1,1:N_i}^{[\mathcal{P}_M]}$ and the knowledge of $\mathcal{P}_M$ without knowing $|\mathcal{P}|$ in advance. This is captured by the following MP-PSI reliability constraint,
\begin{align} \label{MP-PSI Reliability}
\text{[MP-PSI reliability]} \qquad H(\mathcal{P}|Q_{1:M-1,1:N_i}^{[\mathcal{P}_M]}, A_{1:M-1,1:N_i}^{[\mathcal{P}_M]},\mathcal{P}_M)=0
\end{align}

Second, the queries sent by $P_M$ should not leak any information about $\mathcal{P}_M$ except the cardinality of $\mathcal{P}_M$ to any individual database. Thus, $\mathcal{P}_M$ should be independent of all the information available in the $j$th database of $P_i$ for all $ i  \in [1:M-1]$ and $ j \in [1:N_i]$. This is described by the following leader's privacy constraint,
\begin{align} \label{Leader's Privacy}
\text{[Leader's privacy]} \quad I(\mathcal{P}_M;Q_{i,j}^{[\mathcal{P}_M]},A_{i,j}^{[\mathcal{P}_M]},\mathcal{P}_i,\mathcal{R}_{i,j})=0, \quad \forall i  \in [1:M-1], \forall j \in [1:N_i]
\end{align}

\sloppy  Third, client's privacy requires that the leader party does not learn any information other than the intersection $\mathcal{P}$ from the collected answer strings. Let $X_{i,\bar{\cp}}$ be the set of elements in $X_i$ that do not belong to $\cp$, i.e., $X_{i,\bar{\cp}}=\{X_{i,k}: k \in \bar{\cp} \}$.
Hence, the set $\left \{X_{1,\bar{\mathcal{P}}},\cdots,X_{M-1,\bar{\mathcal{P}}} \right\} = \left \{X_{1,k},\cdots,X_{M-1,k}, k \in \bar{\mathcal{P}} \right\}$ should be independent of all the information available in $P_M$. Note that if an element in  $\mathcal{P}_M$ is not in the intersection $\mathcal{P}$, the leader party is supposed to conclude that not all the client parties contain this element simultaneously. On the basis of this fact, we define a new set
$X_{\bar{\mathcal{P}}} = \left \{  \{X_{1,\bar{\mathcal{P}}},\cdots,X_{M-1,\bar{\mathcal{P}}} \}: X_{1,k}+\cdots+X_{M-1,k} < M-1, \forall k \in \mathcal{P}_M \cap \bar{\mathcal{P}}\} \right \}$, we have the following client's privacy constraint,
\begin{align} \label{Client's Privacy}
\text{[Client's privacy]} \qquad I(X_{\bar{\mathcal{P}}};Q_{1:M-1,1:N_i}^{[\mathcal{P}_M]},A_{1:M-1,1:N_i}^{[\mathcal{P}_M]},\mathcal{P}_M) = 0  
\end{align}

For a given field size $K$ and individual parties with associated databases, an MP-PSI achievability scheme is a scheme that satisfies the MP-PSI reliability constraint \eqref{MP-PSI Reliability}, the leader's privacy constraint \eqref{Leader's Privacy} and the client's privacy constraint \eqref{Client's Privacy}. The efficiency of an achievable MP-PSI scheme is measured by its download cost which is the number of downloaded bits (denoted by $D$) by one of the parties in order to compute the intersection $\mathcal{P}$. The optimal download cost is $D^*=\inf D$ over all MP-PSI achievability schemes. 

\section{Main Result}
In this section, we state our main result concerning the performance of our MP-PSI scheme in terms of the download cost. This is summarized in the following theorem, whose proof is given in Section~\ref{Achievable Scheme}.

\begin{theorem} \label{thm1}
	In the MP-PSI problem with $M$ independent parties with data sets $\mathcal{P}_i$, assuming that the parties follow a leader-to-clients communication policy, if the data sets are stored within $N_i$ replicated and non-colluding databases for $i=1, \cdots, M$, then the optimal download cost, $D^*$, is upper bounded by
	\begin{align} \label{Download cost}
D^* \leq \min_{t \in \{1, \cdots, M\}}  \sum_{i \in \{1, \cdots M\}\setminus {t}} \left\lceil \frac{|\cp_t|N_i}{N_i-1}\right\rceil
	\end{align}
\end{theorem}

\begin{remark}
	The minimization problem in (\ref{Download cost}) in Theorem~\ref{thm1} corresponds to the fact that the parties can agree on the party with the minimum $\sum_{i \in \{1, \cdots M\}\setminus {t}} \left\lceil \frac{|\cp_t|N_i}{N_i-1}\right\rceil$ to be the leader party. We note that the leader party may not be the party with the least $|\cp_i|$, as the download cost also depends on the number of the databases at all parties.
\end{remark}

\begin{remark}
     The download cost of our achievability scheme is equal to the sum of the download costs of $M-1$ pair-wise PSI schemes. This implies that there is no penalty incurred due to adopting a stringent clients' privacy constraint over the $E_2$ privacy constraint in \cite{PSIjournal}.   
\end{remark}

\begin{remark}
	 Our achievability scheme is  private in the information-theoretic (absolute) sense and is fairly simple to implement. A drawback of our approach is that it needs multiple replicated non-colluding databases as in the PSI problem in \cite{PSIjournal}; otherwise, our scheme is infeasible if $N_i=1$ for all $i$. 
\end{remark}

\section{Motivating Example: 3 Parties with 3 Databases Each ($M=3$ with $N_1=N_2=N_3=3$)}\label{motivating}
In this section, we motivate our scheme by presenting the following example. In this example, we have $M=3$ parties, each possessing $N_i=3$ replicated and non-colluding databases. Assume that each party stores an independently generated set $\cp_i \subseteq \mathbb{S}_K$, where $\mathbb{S}_K =\{1,2,3,4\}$. Specifically, we assume that $\cp_1=\{1,2\}$, $\cp_2=\{1,3\}$, and $\cp_3=\{1,4\}$. We aim at reliably calculating the intersection $\cp_1 \cap \cp_2 \cap \cp_3 =\{1\}$ without leaking any further information to any of the parties according to the defined communication policy. Without loss of generality, we pick $P_3$ to be the leader party. The remaining parties $P_1$, $P_2$ are referred to as clients.

We map the sets into the corresponding incidence vectors as in \cite{PSIjournal}, i.e., we construct a vector $X_i$, such that $X_{i,k}=1$ if $k \in \cp_i$, hence,
\begin{align}
&\mbox{Party} ~ P_1: \quad \mathcal{P}_1 = \{1,2\} \quad \Rightarrow \quad X_1 = [X_{1,1}\:\: X_{1,2}\:\: X_{1,3} \:\: X_{1,4}]^T = [1 \:\: 1 \:\: 0 \:\:0]^T\\
&\mbox{Party} ~ P_2: \quad \mathcal{P}_2 = \{1,3\} \quad \Rightarrow \quad X_2 = [X_{2,1}\:\: X_{2,2}\:\: X_{2,3} \:\: X_{2,4}]^T = [1 \:\: 0 \:\: 1 \:\:0]^T\\
&\mbox{Party} ~ P_3: \quad \mathcal{P}_3 = \{1,4\} \quad \Rightarrow \quad X_3 = [X_{3,1}\:\: X_{3,2}\:\: X_{3,3} \:\: X_{3,4}]^T = [1 \:\: 0 \:\: 0 \:\:1]^T
\end{align}

To carry out the MP-PSI calculations, the parties agree on a finite field $\mathbb{F}_L$, where $L$ is a prime number such that $L \geq M$. Therefore, we pick $L=3$ in our case, i.e., all summations are performed as modulo-3 arithmetic. 

The leader party $P_3$ initiates the MP-PSI determination protocol by sending queries $Q_{i,j}^{[\cp_3]}$ for $i \in \{1,2\}$ and $j \in \{1,2,3\}$. The queries aim at \emph{privately retrieving} the messages $X_{1,1}$, $X_{1,4}$ and $X_{2,1}$, $X_{2,4}$ using the SPIR retrieval scheme in \cite{SPIR} (the same query structure was introduced in the original work of \cite{PIR_ORI}). Note that in this example we have $N_i=|\cp_3|+1$, thus, the leader party sends exactly $1$ query to each client database. More specifically, let $h_k$, where $k=1, \cdots, 4$, be a random variable picked uniformly and independently from $\mathbb{F}_3$, then, for client party $P_1$, the queries sent from the leader party $P_3$ are generated as follows,
\begin{align}
Q_{1,1}^{[\mathcal{P}_3]} &= [h_1 \:\: h_2 \:\: h_3 \:\: h_4]^T \\
Q_{1,2}^{[\mathcal{P}_3]} &= [h_1+1 \:\: h_2 \:\: h_3 \:\: h_4]^T \\
Q_{1,3}^{[\mathcal{P}_3]} &= [h_1 \:\: h_2 \:\: h_3 \:\: h_4+1]^T
\end{align}
i.e., the leader party sends a random vector $\mathbf{h}=[h_1 \:\: h_2 \:\: h_3 \:\: h_4] \in \mathbb{F}_3^4$ to the first database as a query. The queries for the remaining databases add a $1$ to the positions corresponding to $\cp_3$. For client party $P_2$, the leader party submits the same set of queries,
\begin{align}
Q_{2,1}^{[\mathcal{P}_3]} &= [h_1 \:\: h_2 \:\: h_3 \:\: h_4]^T \\
Q_{2,2}^{[\mathcal{P}_3]} &= [h_1+1 \:\: h_2 \:\: h_3 \:\: h_4]^T \\
Q_{2,3}^{[\mathcal{P}_3]} &= [h_1 \:\: h_2 \:\: h_3 \:\: h_4+1]^T  
\end{align}

Originally in PSI, the client databases obtain the inner product of $X_i$ and $Q_{i,j}^{[\cp_3]}$ and add a common randomness. In MP-PSI, however, we note that applying the answering strategy of \cite{PSIjournal,SPIR} compromises the clients' privacy constraint \eqref{Client's Privacy}. This is due to the fact that the leader, in this case, can decode that $X_{1,4}=0$ and $X_{2,4}=0$ and not only the intersection $\cap_{i=1,2,3} ~ \cp_i$. Consequently, the clients' databases need to share intricate common randomness prior to the retrieval phase to prevent that. To that end, the client parties generate and/or share the following randomness (see Fig.~\ref{MP_PSI_Example1}):

\begin{figure}[ttpb]
	\centering
	\includegraphics[width=1\textwidth]{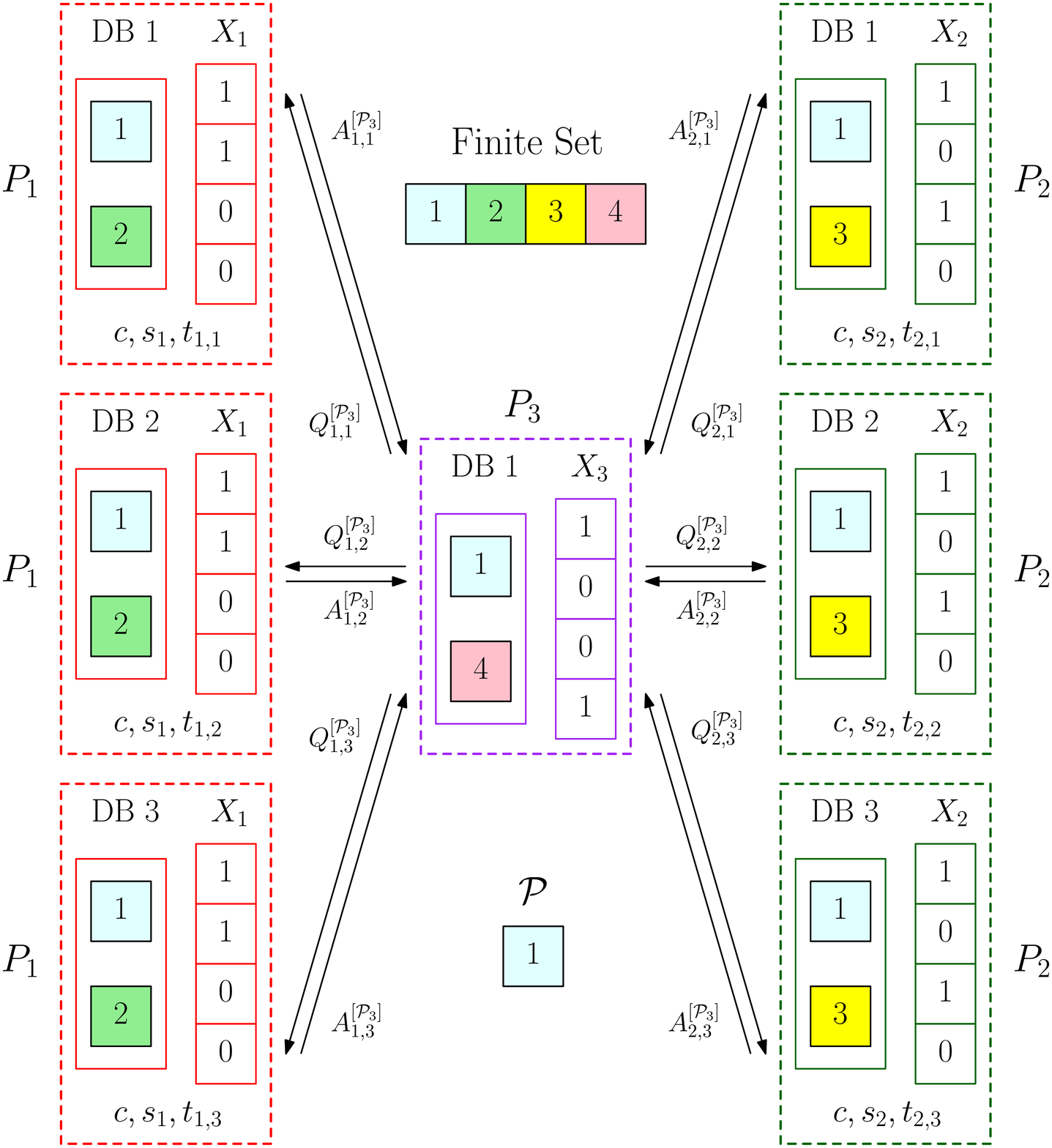}
	\caption{MP-PSI for the motivating example.}
	\label{MP_PSI_Example1}
\end{figure}

\begin{enumerate}
	\item \emph{Local randomness:} This is denoted by the random variable $s_i$, for $i=1,2$. The random variable $s_i$ is picked uniformly from $\mathbb{F}_3$ independent of all data sets and other randomness sources. The local randomness $s_i$ is shared among all the databases belonging to the $i$th client party and not shared with other parties. This local randomness acts as the common randomness needed for SPIR \cite{SPIR}, and is added to the inner product of the incidence vector and the query.
	
	\item \emph{Individual correlated randomness:} This is possessed by each client's database, and is denoted by the random variables $t_{i,j}$ for $i=1,2$, and $j=1,2,3$. This is needed to prevent the leader party from decoding $X_{1,4}$, and $X_{2,4}$. However, since we also need the leader party to decode the intersection, the random variables $t_{i,j}$ need to be correlated such that their effect can be removed if $X_{i,j}$ belongs to the intersection. To that end, we choose $t_{1,1}=t_{2,1}=0$. Database $2$ of the party $P_1$ generates uniformly and independently $t_{1,2}$ from $\mathbb{F}_3$ and sends it to database $2$ of party $P_2$. Database $2$ of the party $P_2$ calculates $t_{2,2}=1-t_{1,2}$. Similarly, database $3$ of the party $P_1$ generates $t_{1,3}$ uniformly and independently from $\mathbb{F}_3$ and shares it with database 3 of $P_2$. Hence,
	\begin{align}
	t_{1,j} \sim \text{uniform}\{0,1,2\}, &\quad j=2,3 \\
	t_{1,j}+t_{2,j}=1, &\quad j=2,3
	\end{align} 
	This randomness is added to each response as well. Note that client parties do not know each other's data sets while generating/sharing this randomness.
	
	\item \emph{Global randomness:} This is denoted by the random variable $c$. The random variable $c$ is generated randomly and independently of all data sets and other randomness variables. The global randomness $c$ is picked uniformly from $\mathbb{F}_3 \setminus \{0\}=\{1,2\}$. The global randomness is shared among all databases of all client parties $P_1$ and $P_2$. The global randomness is used as a multiplier to the responses.
\end{enumerate}

After sharing the common randomness needed to construct the answer strings as shown above, the $j$th database of the $i$th client party responds to the query $Q_{i,j}^{[\cp_3]}$ as follows,
\begin{align}
A_{i,j}^{[\cp_3]}=c(X_i^TQ_{i,j}^{[\cp_3]}+s_i+t_{i,j}), \quad i=1,2, \: j=1,2,3
\end{align}
Hence, noting that $t_{1,1}=0$, the answer strings from $P_1$ can be explicitly written as,
\begin{align}
A_{1,1}^{[\mathcal{P}_3]} &= c\left(\sum\limits_{k=1}^4 h_kX_{1,k} + s_1\right) \\
A_{1,2}^{[\mathcal{P}_3]} &= c\left(\sum\limits_{k=1}^4 h_kX_{1,k} + X_{1,1} + s_1 + t_{1,2}\right) \\
A_{1,3}^{[\mathcal{P}_3]} &= c\left(\sum\limits_{k=1}^4 h_kX_{1,k} + X_{1,4} + s_1 + t_{1,3}\right)
\end{align}
Similarly, the answer strings from $P_2$ are,
\begin{align}
A_{2,1}^{[\mathcal{P}_3]} &= c\left(\sum\limits_{k=1}^4 h_kX_{2,k} + s_2\right) \\
A_{2,2}^{[\mathcal{P}_3]} &= c\left(\sum\limits_{k=1}^4 h_kX_{2,k} + X_{2,1} + s_2 + t_{2,2}\right) \\
A_{2,3}^{[\mathcal{P}_3]} &= c\left(\sum\limits_{k=1}^4 h_kX_{2,k} + X_{2,4} + s_2 + t_{2,3}\right)
\end{align}

Note that, by this construction, the local randomness $s_i$ is used to protect the random sum $\sum_{k=1}^4 h_kX_{i,k}$ as in SPIR, and the individual randomness $t_{i,j}$ is needed to prevent the leader party from directly decoding $X_{i,j+1}$. Note that $s_1$ and $s_2$ need to be independent to avoid the information leakage about the relationship between $\sum_{k=1}^4 h_kX_{1,k}$ and $\sum_{k=1}^4 h_kX_{2,k}$. 

\paragraph{Reliability:} To calculate $\cap_{i=1,2,3} ~ \cp_i$ based on the answer strings the leader party has received, the leader party subtracts $A_{1,1}^{[\mathcal{P}_3]}$ and $A_{2,1}^{[\mathcal{P}_3]}$ from the remaining answer strings. Denote the result of subtraction related to the $j$th element in $\mathbb{S}_K$ at $P_i$ by $Z_{i,j}$. This leads to, 
\begin{align} 
Z_{1,1}=c(X_{1,1} + t_{1,2}) &= A_{1,2}^{[\mathcal{P}_3]} - A_{1,1}^{[\mathcal{P}_3]} \label{Z1}\\
Z_{1,4}=c(X_{1,4} + t_{1,3}) &= A_{1,3}^{[\mathcal{P}_3]} - A_{1,1}^{[\mathcal{P}_3]} \\
Z_{2,1}=c(X_{2,1} + t_{2,2}) &= A_{2,2}^{[\mathcal{P}_3]} - A_{2,1}^{[\mathcal{P}_3]} \\
Z_{2,4}=c(X_{2,4} + t_{2,3}) &= A_{2,3}^{[\mathcal{P}_3]} - A_{2,1}^{[\mathcal{P}_3]}  \label{Z4}
\end{align}

Now, let $E_j$ be an indicator of having the $j$th element in $\mathbb{S}_K$ in the intersection $\cap_{i=1,2,3} ~ \cp_i$, such that $E_j=0$ if and only if $j \in \cap_{i=1,2,3} ~ \cp_i$. To that end, define $E_j$ as the modulo-$L$ sum of $Z_{i,j}$ along all clients, i.e.,
\begin{align}
E_j=\sum_{i=1}^{M-1} Z_{i,j}
\end{align}

Looking deeper at $E_1$, we note that,
\begin{align}
E_1 &=Z_{1,1}+Z_{2,1} \\
    &=c(X_{1,1}+X_{2,1}+t_{1,2}+t_{2,2}) \\
    &=c(X_{1,1}+X_{2,1}+1)
\end{align}
where $t_{1,2}+t_{2,2}=1$ by the construction of the individual correlated randomness. Therefore, $E_1=0$ if and only if $X_{1,1}=1$ and $X_{2,1}=1$ simultaneously. In this case, $E_1=0$ irrespective of the value of $c$ and the leader party verifies that $\{1\} \subseteq \cap_{i=1,2,3} ~ \cp_i$.

On the other hand, when $P_3$ calculates $E_4$,
\begin{align}
E_4=Z_{1,4}+Z_{2,4}=c(X_{1,1}+X_{2,1}+1) \neq 0
\end{align}

Consequently, the leader party confirms that $\cap_{i=1,2,3} ~ \cp_i=\{1\}$ and does not include $4$.

\paragraph{Leader's Privacy:} The leader's privacy constraint follows from the user's privacy constraint of the inherent SPIR scheme \cite{SPIR}. The queries of the leader to any party have the same structure as the queries of the user in the SPIR problem. More specifically, the privacy of leader party is preserved as each element in the queries is uniformly distributed over the finite field $\mathbb{F}_3$. Hence, no information about $\mathcal{P}_3$ is leaked from the queries.

\paragraph{Client's Privacy:} To see the client's privacy, we note that no information is leaked about $\overline{\cp_1 \cap \cp_3}$ or $\overline{\cp_2 \cap \cp_3}$ due to $s_1$ and $s_2$, respectively. Nevertheless, in MP-PSI, we need to verify that the leader does not know which of the two parties possesses the element $\{4\}$, i.e., knowing the fact that $E_4 \neq 0$, we need to show that  $\mathbb{P}(X_{1,4}+X_{2,4}=0)=\mathbb{P}(X_{1,4}+X_{2,4}=1)=\frac{1}{2}$. Specifically, if $E_4$ is $1$, $\mathbb{P}(X_{1,4} + X_{2,4} = 0) = \mathbb{P}(X_{1,4} + X_{2,4} = 1) = \frac{1}{2}$ because $c$ is uniformly distributed over $1$ and $2$ and the sum $t_{1,3} + t_{2,3}=1$ by construction. The conclusion is exactly the same when $E_4$ equals $2$. Thus, the only information that $P_3$ can obtain for the element $4$ is that client parties $P_1$ and $P_2$ do not contain it at the same time (this is no further leak, as if they did contain it at the same time, it would have been in the intersection). Hence, $c$ is used such that the leader party $P_3$ does not know whether the sum $X_{1,4} + X_{2,4}$ is $0$ or $1$.   

\paragraph{Download Cost:} In our example, the leader party $P_3$ downloads $N_i=|\cp_M|+1$ symbols from each client party. Hence, the total download cost is $D=(M-1)(|\cp_M|+1)=6$.
  
\section{Achievability Proof}\label{Achievable Scheme}
In this section, we describe our general achievable scheme for MP-PSI for arbitrary number of parties $M$, arbitrary set sizes $|\cp_i|$, and arbitrary number of databases per party $N_i$, for $i \in \{1, \cdots, M\}$. The leader's querying policy is based on the SPIR scheme presented in \cite{SPIR} (originally introduced in \cite{PIR_ORI}). Our novel ideas in this scheme are concerned with the construction of the answering strings. More specifically, the scheme hinges on the intricate design of generating and sharing common randomness among the clients' databases in such a way that the leader party cannot learn anything but the intersection $\cap_{i=1}^M \cp_i$.  

\subsection{General Achievability Scheme}
In the following, assume that $\cp_i \subseteq \mathbb{S}_K$, where $|\mathbb{S}_K|=K$.

\begin{enumerate}
	\item \emph{Initialization:} The parties agree on a retrieval finite field $\mathbb{F}_L$ to carry out the calculations needed for MP-PSI determination protocol. $L$ is chosen such that,
	\begin{align}
	 L=\min\:\{L \geq M: L \:\text{is a prime}\}
	\end{align}
	The parties agree on a leader $P_{t^*}$ such that:
	\begin{align}
	t^*= \arg \min_{t \in \{1, \cdots, M\}} \sum_{i \neq t} \left\lceil \frac{|\cp_t|N_i}{N_i-1}\right\rceil
	\end{align}
	Without loss of generality, we assume that $t^*=M$ in the sequel. Furthermore, assume that $\cp_{t^*}=\cp_M=\{Y_1, Y_2, \cdots, Y_R\}$ with cardinality $|\cp_M|=R$.
	
	\item \emph{Query generation:} The leader party $P_M$ independently and uniformly generates $\kappa$ random vectors $\{\mathbf{h}_1, \mathbf{h}_2, \cdots, \mathbf{h}_\kappa\}$, where $\kappa$ is given by,
	\begin{align}
	\kappa = \max_{i \in \{1, \cdots, M-1\}}  \left\lceil \frac{|\cp_M|}{N_i-1}\right\rceil
	\end{align}	
	The vector $\mathbf{h}_\ell$, for $\ell=1,2, \cdots, \kappa$ is picked uniformly from $\mathbf{F}_L^K$ such that,
	\begin{align}
		\mathbf{h}_\ell=[h_\ell(1) \quad h_\ell(2) \quad \cdots \quad h_\ell(K)]
	\end{align}
	Denote $\eta_i=\left\lceil \frac{|\cp_M|}{N_i-1}\right\rceil$, and let $\cp_M^{\ell_i}=\{Y_1^{\ell_i}, Y_2^{\ell_i}, \cdots Y_{N_i-1}^{\ell_i}\}$, for $i=1, \cdots, M-1$.
	The leader party $P_M$ submits $\eta_i$ random vectors from $\{\mathbf{h}_1, \mathbf{h}_2, \cdots, \mathbf{h}_\kappa\}$ to the first database of the $i$th client party as queries. Each submitted random vector can be reused in the remaining $N_i-1$ databases to retrieve $N_i-1$ symbols. This can be done by adding 1 to the positions corresponding to the desired symbols. More specifically, assume that $\cp_M=\cup_{\ell_i=1}^{\eta_i} \cp_M^{\ell_i}$, where $\cp_M^{\ell_i} \subseteq \cp_M$ are disjoint partitions of $\cp_M$ such that $|\cp_M^{\ell_i}|=N_i-1$ (except potentially for the last subset $\cp_M^{\eta_i}$), then for $i=1, 2, \cdots, M-1$, the query structure is given by:
	\begin{align}
	Q_{i,1}^{[\mathcal{P}_M^{\ell_1}]} &= [h_1(1)\quad h_1(2) \quad \cdots \quad h_1(K)\} \\
	Q_{i,2}^{[\mathcal{P}_M^{\ell_1}]} &= [h_1(1)\quad \cdots\quad h_1(Y_1^{\ell_1}-1)\quad h_1(Y_1^{\ell_1})+1 \quad h_1(Y_1^{\ell_1}+1)\quad \cdots\quad h_1(K)] \\
	&\vdots \notag\\
	Q_{i,N_i}^{[\mathcal{P}_M^{\ell_1}]} &= [h_1(1)\quad \cdots\quad h_1(Y_{N_i-1}^{\ell_1}-1)\quad h_1(Y_{N_i-1}^{\ell_1})+1 \quad h_1(Y_{N_i-1}^{\ell_1}+1)\quad \cdots\quad h_1(K)] \\
	&\vdots \notag\\
	Q_{i,1}^{[\mathcal{P}_M^{\eta_i}]} &= [h_{\eta_i}(1)\quad h_{\eta_i}(2) \quad \cdots \quad h_{\eta_i}(K)\} \\
	Q_{i,2}^{[\mathcal{P}_M^{\ell_1}]} &= [h_{\eta_i}(1)\quad \cdots\quad h_{\eta_i}(Y_1^{\eta_i}-1)\quad h_{\eta_i}(Y_1^{\eta_i})+1 \quad h_{\eta_i}(Y_1^{\eta_i}+1)\quad \cdots\quad h_{\eta_i}(K)] \\
	&\vdots \notag\\
	Q_{i,N_i}^{[\mathcal{P}_M^{\ell_1}]} &= [h_{\eta_i}(1)\quad \cdots\quad h_{\eta_i}(Y_{N_i-1}^{\eta_i}-1)\quad h_{\eta_i}(Y_{N_i-1}^{\eta_i})+1 \quad h_{\eta_i}(Y_{N_i-1}^{\eta_i}+1)\quad \cdots\quad h_{\eta_i}(K)]
	\end{align}
	i.e., $P_M$ simply partitions the set $\cp_M$ into subsets of size $N_i-1$. For each set, $P_M$ uses different $\mathbf{h}_\ell$. $P_M$ submits $\mathbf{h}_\ell$ into the first database. For the remaining databases, it adds 1 for the positions that corresponds to the partition.
	
	\item \emph{Common randomness generation:} In order to respond to the leader party, the clients need to generate and share common randomness. Specifically, there are three types of randomness:
	
	\begin{itemize} 
		\item \emph{Local randomness:} This is denoted by $\mathbf{s}_i=[s_i(1) \:\: s_i(2) \:\: s_i(\eta_i)]$. Each element of $\mathbf{s}_i$ is generated independently and uniformly from $\mathbb{F}_L$. The local randomness $\mathbf{s}_i$ is shared between the databases associated with $P_i$. The local randomness is added to the responses as in SPIR \cite{SPIR}. Note that each database uses a different element from $\mathbf{s}_i$ for each submitted query.
		
		\item \emph{Individual correlated randomness:} The $j$th database associated with the $i$th client possesses an individual randomness $\mathbf{t}_{i,j}=[t_{i,j}(1) \:\: t_{i,j}(2) \:\: t_{i,j}(\eta_i)]$ for $i=1, \cdots, M-1$, and $j=1, \cdots, N_i$. The elements $t_{i,1}=0$ for all $i$. For $i=1, \cdots, M-2$, the vector $\mathbf{t}_{i,j}$ is independently and uniformly picked from $\mathbf{F}_L^{\eta_i}$. All these random vectors are sent to the party $P_{M-1}$. The client $P_{M-1}$ generates its individual randomness $\mathbf{t}_{M-1,j}$ according to the received individual randomness from the remaining parties. For simplicity, let us (re)denote the individual randomness components by $\tilde{t}_{i,k}$, where $i$ is the index of the client party and $k=1,2, \cdots, R$ is just a monotonically increasing index of the randomness component used within the databases $2$ to $N_i$ of the $i$th client. Thus,
		\begin{align}
		\tilde{t}_{i,1}=t_{i,2}(1), \quad \tilde{t}_{i,1}=t_{i,2}(2), \cdots, \tilde{t}_{i,R}=t_{i,N_i}(\eta_i)
		\end{align}
		With this re-definition, the client $P_{M-1}$ calculates its individual randomness as,
		\begin{align}
		\tilde{t}_{M-1,j}=L-(M-1)-\sum_{i=1}^{M-2} \tilde{t}_{i,j}, \quad j=1, 2,\cdots, R
		\end{align}
		This ensures that the individual randomness are correlated such that $\sum_{i=1}^{M-1} \tilde{t}_{i,j}=L-(M-1)$. The individual randomness is added to the responses.  
		
		\item \emph{Global randomness:} This is denoted by $c$. $c$ is picked uniformly and independently from $\mathbb{F}_L\setminus \{0\}$. $c$ is shared among all the databases at all clients. $c$ is used as a multiplier for the answering string.
	\end{itemize}
	
\item \emph{Response generation:} The clients respond to the submitted queries by using the queries as a combining vector to their contents, i.e., each database calculates the inner product of the query and its contents. Next, it adds the local and individual randomness. Finally, it multiplies the result by the global randomness. More specifically, the answer string of the $j$th database, which is associated with the $i$th client to retrieve one of the elements of the partition $\cp_M^{\ell_i}$, $A_{i,j}^{[\cp_M^{\ell_i}]}$, is given by,
\begin{align}
A_{i,j}^{[\cp_M^{\ell_i}]}=c\left(X_i^T Q_{i,j}^{[\cp_M^{\ell_i}]}+s_i(\ell_i)+t_{i,j}(\ell_i)\right)
\end{align}    
\end{enumerate}
From the collected answers the leader party can determine the intersection $\cap_{i=1}^M \cp_i$ reliably and privately. 

\subsection{Download Cost, Reliability, Leader's Privacy, Clients' Privacy}

\paragraph{Download cost:} By observing the queries associated with the MP-PSI scheme in the previous section, one can note that the desired symbols are divided into $\eta_i=\left\lceil \frac{|\cp_M|}{N_i-1}\right\rceil$ subsets. Each subset consists of $N_i-1$ desired symbols. The leader needs to download 1 bit from all $N_i$ databases to query the entire subset, as the leader downloads useless random linear combination of the contents from the first database. Hence, the download cost is given by,
\begin{align}
D&= \sum_{i=1}^{M-1} N_i \eta_i \\
 &=\sum_{i=1}^{M-1} \left\lceil \frac{|\cp_M|N_i}{N_i-1}\right\rceil
\end{align}
\paragraph{Reliability:} To verify reliability, we follow the leader's processing of the responses. First, we note that the answer string that is returned from database 1 is a random linear combination of the contents of the database besides the common randomness, and is given by,
\begin{align}
	A_{i,1}^{[\mathcal{P}_M^{\ell_i}]} &= c\left(\sum\limits_{k=1}^K h_{\ell_i}(k)X_{i,k} + s_i(\ell_i)\right), \quad i=1, \cdots, M-1
\end{align}
Note that $t_{i,1}=0$ by construction. The leader subtracts this response from each response that belongs to the same partition. Denote the subtraction result at the $i$th client that contains the element $X_{i,k}$ by $Z_{i,k}$, hence,
\begin{align}
Z_{i,k}=c(X_{i,k}+\tilde{t}_{i,k})=A_{i,j^*}^{[\cp_M^{\ell_i}]}-A_{i,1}^{[\cp_M^{\ell_i}]}, \quad k \in \cp_M^{\ell_i} 
\end{align}
for some unique $j^*$ that $A_{i,j^*}^{[\cp_M^{\ell_i}]}$ is a response of the query that adds 1 to the $k$th position of the query vector. In particular, for the special case of $N_i=|\cp_i|+1$ for all $i=1, \cdots, M-1$, we have $j^*=k+1$ and $\cp_M^{\ell_i}=\cp_M$ (one partition). Note that we used the alternative notation $\tilde{t}_{i,k}$ as it is counted in sequence. 

Next, the leader constructs the intersection indicator variable $E_k$, where $E_k$ is given by,
\begin{align}
E_k&=\sum_{i=1}^{M-1} Z_{i,k} \\
   &=c\left(\sum_{i=1}^{M-1} X_{i,k}+\sum_{i=1}^{M-1} \tilde{t}_{i,k}\right)\\
   &=c\left(\sum_{i=1}^{M-1} X_{i,k}+L-(M-1)\right) \label{Ek}
\end{align} 
where \eqref{Ek} follows from the construction of the individual randomness. Now, the element $E_k=0$ if and only if $\sum_{i=1}^{M-1} X_{i,k}=M-1$, which implies that $X_{i,k}=1$ for all $i=1,2, \cdots, M-1$. Consequently, $Y_k \in \cap_{i=1}^M \cp_i$ if and only if $E_k=0$. This proves the reliability of the scheme.  

\paragraph{Leader's privacy:} The leader's privacy follows from the fact that the random vectors $\{\mathbf{h}_1, \cdots, \mathbf{h}_\kappa\}$ are uniformly generated over $\mathbb{F}_L^K$. Adding $1$ to these vectors does not change the statistical distribution of the vector. Since the leader submits independent vectors each time it queries a database, all queries are equally likely and the leader's privacy is preserved.

\paragraph{Clients' privacy:}  
Without loss of generality, we derive the proof of the client's privacy for the homogeneous number of databases, i.e., $N_i = R+1, \forall i \in [1:M-1]$. The general proof in the heterogeneous case follows the same steps and after removing the response of the first databases, we will be left with $Z_{i,k}$ that has the same structure of homogeneous case. Consequently, we present the homogeneous case here for convenience only. In the following proof, we adopt the notation that for a random variable $\zeta_{i,j}$ indexed by two indices $(i,j)$,
\begin{align}
    \zeta_{i_1:i_M,j_1:j_R}=\{\zeta_{i,j}: i \in \{i_1, \cdots, i_M\}, \: j \in \{j_1, \cdots, j_R\}\}
\end{align}

For the proof, we need the following lemmas. Lemma~\ref{lemma1} shows that the effect of the local randomness is to make the response of the first database at all parties independent of $X_{\bar{\cp}}$.

\begin{lemma}\label{lemma1}
For the presented achievable scheme, we have,
\begin{align}
    I(X_{\bar{\mathcal{P}}};A_{1:M-1,1}^{[\mathcal{P}_M]}|Z_{1:M-1,Y_1:Y_R},Q_{1:M-1,1:N_i}^{[\mathcal{P}_M]},\mathcal{P}_M)=0
\end{align}
\end{lemma}

\begin{Proof}
Intuitively, the proof follows from the fact that $A_{i,1}^{[\mathcal{P}_M]}, i \in [1:M-1]$ is a random variable uniformly distributed over $[0:L-1]$ because of the local randomness $s_i$, and thus, is independent of the data sets, queries and the subtraction results. More specifically, 
\begin{align}
    &I(X_{\bar{\mathcal{P}}};A_{1:M-1,1}^{[\mathcal{P}_M]}|Z_{1:M-1,Y_1:Y_R},Q_{1:M-1,1:N_i}^{[\mathcal{P}_M]},\mathcal{P}_M)\notag\\
    &=H(A_{1:M-1,1}^{[\mathcal{P}_M]}|Z_{1:M-1,Y_1:Y_R},Q_{1:M-1,1:N_i}^{[\mathcal{P}_M]},\mathcal{P}_M)-H(A_{1:M-1,1}^{[\mathcal{P}_M]}|X_{\bar{\mathcal{P}}},Z_{1:M-1,Y_1:Y_R},Q_{1:M-1,1:N_i}^{[\mathcal{P}_M]},\mathcal{P}_M)\\
    &\leq H(A_{1:M-1,1}^{[\cp_M]})-H(A_{1:M-1,1}^{[\mathcal{P}_M]}|X_{1:M-1},c,X_{\bar{\mathcal{P}}},Z_{1:M-1,Y_1:Y_R},Q_{1:M-1,1:N_i}^{[\mathcal{P}_M]},\mathcal{P}_M)\\
    &\leq (M-1)-H(s_1,\cdots, s_{M-1})\\
    &=(M-1)-(M-1)=0 
\end{align}
This concludes the proof, since $I(X_{\bar{\mathcal{P}}};A_{1:M-1,1}^{[\mathcal{P}_M]}|Z_{1:M-1,Y_1:Y_R},Q_{1:M-1,1:N_i}^{[\mathcal{P}_M]},\mathcal{P}_M) \geq 0$.
\end{Proof}

Lemma~\ref{lemma2} asserts that for $i \in [1:M-2]$, $j \in [1:R]$ the effect of individual randomness $t_{i,j+1}$ is to force the random variables $Z_{i,Y_j}$ to be independent of $X_{\bar{\cp}}$. Note that we do not claim anything about $Z_{M-1,Y_j}$ as the individual randomness are correlated at party $M-1$.

\begin{lemma}\label{lemma2}
    For the presented scheme, we have,
    \begin{align}
        I(X_{\bar{\cp}};Z_{1:M-2,Y_1:Y_R}|E_{Y_1:Y_R},Q_{1:M-1,1:N_i}^{[\cp_M]},\cp_M)=0
    \end{align}
\end{lemma}

\begin{Proof}
Intuitively, similar to the proof of Lemma~\ref{lemma1}, the proof follows from the fact that $Z_{i,Y_j}, i \in [1:M-2], j \in [1:R]$ is a random variable uniformly distributed over $[0:L-1]$ because of the individual randomness $t_{i,j+1}$, and thus, is independent of the data sets, queries, and the data sets in the client parties $E_{Y_j}$,
\begin{align}
   &I(X_{\bar{\cp}};Z_{1:M-2,Y_1:Y_R}|E_{Y_1:Y_R},Q_{1:M-1,1:N_i}^{[\cp_M]},\cp_M)\notag\\
   &=H(Z_{1:M-2,Y_1:Y_R}|E_{Y_1:Y_R},Q_{1:M-1,1:N_i}^{[\cp_M]},\cp_M)-H(Z_{1:M-2,Y_1:Y_R}|X_{\bar{\cp}},E_{Y_1:Y_R},Q_{1:M-1,1:N_i}^{[\cp_M]},\cp_M)\\
   &\leq H(Z_{1:M-2,Y_1:Y_R})-H(Z_{1:M-2,Y_1:Y_R}|X_{1:M-1},c,X_{\bar{\cp}},E_{Y_1:Y_R},Q_{1:M-1,1:N_i}^{[\cp_M]},\cp_M)\\
   &\leq ((M-2)R)-H(t_{1:M-2,Y_1:Y_R})\\
   &=((M-2)R)-((M-2)R)=0
\end{align}
This concludes the proof as the reverse implication is true by the non-negativity of mutual information.
\end{Proof}

The following lemma asserts that indicator functions $E_{Y_j}$ for all $j$ do not leak any information about $X_{\bar{\cp}}$.

\begin{lemma}\label{lemma3}
For the presented scheme, we have,
\begin{align}
     I(X_{\bar{\mathcal{P}}};E_{Y_1:Y_R},Q_{1:M-1,1:N_i}^{[\mathcal{P}_M]},\cp_M)=0
\end{align}
\end{lemma}

\begin{Proof}
Note that if $Y_j \in \mathcal{P}_M$ is in the intersection, $E_{Y_j} = 0$ has nothing to do with $X_{\bar{\mathcal{P}}}$ since $X_{\bar{\mathcal{P}}}$ is defined on the elements not in the intersection. However, if $Y_j$ is not in the intersection, $E_{Y_j} = c(X_{1,Y_j}+\cdots+X_{M-1,Y_j}+L-(M-1)), Y_j \in \mathcal{P}_M \cap \bar{\mathcal{P}}$ received by the leader party would be a realization within the range of $\mathbb{F}_L\setminus \{0\}$ because of the global randomness $c$. However, the leader party only knows that the global randomness $c$ is uniformly distributed over $\mathbb{F}_L\setminus \{0\}$ and has no information about the specific value of $c$ in the client parties. As a result, from the perspective of the leader part $P_M$, $X_{1,Y_j}+\cdots+X_{M-1,Y_j}+L-(M-1)$ is uniformly distributed over $[1:L-1]$ according to the information contained in $E_{Y_j}$. This comes from the fact that the set $\mathbb{F}_L\setminus \{0\}$ of all $L-1$ non-zero elements must form a finite cyclic group under multiplication given a finite field $\mathbb{F}_L$. That means that, in the additive table under multiplication operation, each element in $\mathbb{F}_L\setminus \{0\}$ appears precisely once in each row and column of the table. The probability $\mathbb{P}(X_{1,Y_j}+\cdots+X_{M-1,Y_j}+L-(M-1) = l)$ would always be $\frac{1}{L-1}$ for any $l \in [1:L-1]$. Then, $X_{1,Y_j}+\cdots+X_{M-1,Y_j}$ is uniformly distributed over $[M-L:M-2]$ (i.e., $[0:M-2] \cup [M:L-1]$) and we can further conclude that $X_{1,Y_j}+\cdots+X_{M-1,Y_j}$ is uniformly distributed over $[0:M-2]$ because its largest possible value is $M-2$ if $Y_j$ is not in the intersection. Thus, the only information we can learn from $E_{Y_1},\cdots,E_{Y_R}$ and the accompanying queries about $X_{\bar{\mathcal{P}}}$ is $X_{1,k}+\cdots+X_{M-1,k} < M-1, \forall k \in \mathcal{P}_M \cap \bar{\mathcal{P}}$ without knowing the specific value of $X_{1,k}+\cdots+X_{M-1,k}$, which already exists in the definition of $X_{\bar{\mathcal{P}}}$. Thus, we obtain,
 \begin{align}
     I(X_{\bar{\mathcal{P}}};E_{Y_1:Y_R},Q_{1:M-1,1:N_i}^{[\mathcal{P}_M]},\cp_M) 
     &=I(X_{\bar{\mathcal{P}}};E_{Y_1:Y_R}|Q_{1:M-1,1:N_i}^{[\mathcal{P}_M]},\cp_M) \label{lemma3.1}\\
     &=H(X_{\bar{\mathcal{P}}}|Q_{1:M-1,1:N_i}^{[\mathcal{P}_M]},\cp_M)-H(X_{\bar{\mathcal{P}}}|E_{Y_1:Y_R},Q_{1:M-1,1:N_i}^{[\mathcal{P}_M]},\cp_M)\\
     &=H(X_{\bar{\mathcal{P}}})-H(X_{\bar{\mathcal{P}}}) \\ 
     &=0
\end{align}
where \eqref{lemma3.1} follows from the fact that queries and $\mathcal{P}_M$ are independent of the data sets in the client parties $E_{Y_j}$ in \eqref{Independence}. 
\end{Proof}

Now, we are ready to show that our achievability satisfies the client's privacy constraint,
\begin{align}
     &I(X_{\bar{\mathcal{P}}};Q_{1:M-1,1:N_i}^{[\mathcal{P}_M]},A_{1:M-1,1:N_i}^{[\mathcal{P}_M]},\mathcal{P}_M) \notag\\
     &=I(X_{\bar{\mathcal{P}}};A_{1:M-1,1}^{[\cp_M]},Z_{1:M-1,Y_1:Y_R},Q_{1:M-1,1:N_i}^{[\mathcal{P}_M]},\mathcal{P}_M) \label{client1}\\
     &=I(X_{\bar{\mathcal{P}}};Z_{1:M-1,Y_1:Y_R},Q_{1:M-1,1:N_i}^{[\mathcal{P}_M]},\mathcal{P}_M)\!+\!I(X_{\bar{\mathcal{P}}};A_{1:M-1,1}^{[\cp_M]}|Z_{1:M-1,Y_1:Y_R},Q_{1:M-1,1:N_i}^{[\mathcal{P}_M]},\mathcal{P}_M)\\
     &=I(X_{\bar{\mathcal{P}}};Z_{1:M-1,Y_1:Y_R},Q_{1:M-1,1:N_i}^{[\mathcal{P}_M]},\mathcal{P}_M)\label{client2}\\
     &=I(X_{\bar{\mathcal{P}}};Z_{1:M-2,Y_1:Y_R},E_{Y_1:Y_R},Q_{1:M-1,1:N_i}^{[\mathcal{P}_M]},\mathcal{P}_M) \label{client3}\\
     &=I(X_{\bar{\mathcal{P}}};E_{Y_1:Y_R},Q_{1:M-1,1:N_i}^{[\mathcal{P}_M]},\mathcal{P}_M)+I(X_{\bar{\mathcal{P}}};Z_{1:M-2,Y_1:Y_R}|E_{Y_1:Y_R},Q_{1:M-1,1:N_i}^{[\mathcal{P}_M]},\mathcal{P}_M)\\
     &=I(X_{\bar{\mathcal{P}}};E_{Y_1:Y_R},Q_{1:M-1,1:N_i}^{[\mathcal{P}_M]},\mathcal{P}_M) \label{client4}\\
     &=0 \label{client5}
\end{align}
where \eqref{client1} follows from the fact that there is a bijective transformation between $A_{1:M-1,1:N_i}^{[\cp_M]}$ and $(A_{1:M-1,1}^{[\cp_M]},Z_{1:M-1,Y_1:Y_R})$, \eqref{client2} follows from Lemma~\ref{lemma1}, \eqref{client3} follows from the fact that there is a bijective transformation between $Z_{1:M-1,Y_1:Y_R}$ and $(Z_{1:M-2,Y_1:Y_R},E_{Y_1:Y_R})$, \eqref{client4} follows from Lemma~\ref{lemma2}, and \eqref{client5} follows from Lemma~\ref{lemma3}.

\section{Further Examples}
In this section, we present two examples of our achievable scheme. Unlike the motivating example in Section~\ref{motivating}, in these examples, the number of databases per party does not need to be $N_i=|\cp_M|+1$ or even be homogeneous in general\footnote{For $N_i>|\cp_M|+1$, we just use any arbitrary $|\cp_M|+1$ databases to execute the MP-PSI determination protocol.}.

\subsection{An Example for $N_i<|\cp_M|+1$} 
In this example, we use the same setting of Section~\ref{motivating} with $\cp_1=\{1,2\}$, $\cp_2=\{1,3\}$, and $\cp_3=\{1,4\}$ with $P_3$ being the leader party and the retrieval field being $\mathbb{F}_3$. The incidence vectors $X_i$, for $i=1,2$ remain the same. However, to illustrate that our scheme works for $N_i<|\cp_M|+1$, we assume that $N_1=N_2=2$. As we will show next, when $N_i<|\cp_M|+1$, we need to send $\kappa=\eta_i=\left\lceil\frac{|\cp_M|}{N_i-1}\right\rceil=2$ queries to the first database of the $i$th party (in contrast to 1 query only when $N_i\geq|\cp_M|+1$). Moreover, the common randomness components $\mathbf{s}_i$, and $\mathbf{t}_{i,j}$ need to be vectors of size $\left\lceil\frac{|\cp_M|}{N_i-1}\right\rceil=2$. Note that, in this case, the leader's set is divided into 2 subsets $\cp_M^{\ell_1}=\{1\}$ and $\cp_M^{\ell_1}=\{4\}$ as $|\cp_M^{\ell_i}|=N_i-1=1$.

For the queries, since both client parties have the same number of databases, the leader $P_3$ submits the same query vectors to the databases of both clients. The first databases of each client receives 2 uniformly generated vectors $\mathbf{h}, \bar{\mathbf{h}} \in \mathbb{F}_3^4$, where $\mathbf{h}=[h_1 \: h_2 \: h_3 \: h_4]^T$ and $\bar{\mathbf{h}}=[\bar{h}_1 \: \bar{h}_2 \: \bar{h}_3 \: \bar{h}_4]^T$. $P_3$ submits the same two vectors to the second databases of $P_1$ and $P_2$ with adding 1 to the desired positions. More specifically, let $Q_{i,j}^{[k]}$ be the query to the $j$th database of $P_i$ to retrieve the element $k$, then $P_3$ submits the following queries:
\begin{align}
Q_{1,1}^{[1]}&=Q_{2,1}^{[1]}=[h_1 \:\: h_2 \:\: h_3 \:\: h_4]^T \\
Q_{1,2}^{[1]}&=Q_{2,2}^{[1]}=[h_1+1 \:\: h_2 \:\: h_3 \:\: h_4]^T \\
Q_{1,1}^{[4]}&=Q_{2,1}^{[4]}=[\bar{h}_1 \:\: \bar{h}_2 \:\: \bar{h}_3 \:\: \bar{h}_4]^T \\
Q_{1,2}^{[4]}&=Q_{2,2}^{[4]}=[\bar{h}_1 \:\: \bar{h}_2 \:\: \bar{h}_3 \:\: \bar{h}_4+1]^T
\end{align}  

At the clients' side, the clients share a global randomness $c \sim \text{uniform}\{1,2\}$ among all the databases of both clients. For $i=1,2$, the $i$th client generates and shares a local randomness $\mathbf{s}_i=[s_i(1) \:\: s_i(2)]^T$, such that $s_i(\ell) \sim \text{uniform}\{0,1,2\}$ among the databases that belong to the $i$th client. Finally, for $i=1,2$, the second database of the $i$th client has an individual correlated randomness $\mathbf{t}_{i,2}=[t_{i,2}(1) \:\: t_{i,2}(2)]^T$, such that $t_{1,2}(1) \sim t_{1,2}(2) \sim \text{uniform}\{0,1,2\}$, $t_{1,2}(1)+t_{2,2}(1)=1$, and $t_{1,2}(2)+t_{2,2}(2)=1$. Assume that $t_{1,1}=t_{2,1}=0$. All randomness components are independently generated of each other and of the data sets.

The answer string $A_{i,j}^{[k]}$, for $i=1,2$, $j=1,2$, $k=1,4$, is given by, 
\begin{align}
A_{i,j}^{[k]}=c\left(X_i^T Q_{i,j}^{[k]}+s_i(\ell(k))+t_{i,j}(\ell(k))\right)
\end{align}
where $\ell(1)=1$ and $\ell(4)=2$.

Thus, the leader party receives the following answer strings from $P_1$,
\begin{align}
A_{1,1}^{[1]} &= c\left(\sum\limits_{k=1}^4 h_kX_{1,k} + s_1(1)\right) \\
A_{1,2}^{[1]} &= c\left(\sum\limits_{k=1}^4 h_kX_{1,k} + X_{1,1} + s_1(1) + t_{1,2}(1)\right) \\
A_{1,1}^{[4]} &= c\left(\sum\limits_{k=1}^4 \bar{h}_kX_{1,k} + s_1(2)\right) \\
A_{1,2}^{[4]} &= c\left(\sum\limits_{k=1}^4 \bar{h}_kX_{1,k} + X_{1,4} + s_1(2) + t_{1,2}(2)\right) 
\end{align}
and the following answer strings from $P_2$,
\begin{align}
A_{2,1}^{[1]} &= c\left(\sum\limits_{k=1}^4 h_kX_{2,k} + s_2(1)\right) \\
A_{2,2}^{[1]} &= c\left(\sum\limits_{k=1}^4 h_kX_{2,k} + X_{1,1} + s_2(1) + t_{2,2}(1)\right) \\
A_{2,1}^{[4]} &= c\left(\sum\limits_{k=1}^4 \bar{h}_kX_{2,k} + s_2(2)\right) \\
A_{2,2}^{[4]} &= c\left(\sum\limits_{k=1}^4 \bar{h}_kX_{2,k} + X_{1,4} + s_2(2) + t_{2,2}(2)\right) 
\end{align}

The leader party constructs the subtractions $Z_{i,j}$ as follows,
\begin{align}
Z_{1,1}=c(X_{1,1}+t_{1,2}(1))=A_{1,2}^{[1]}-A_{1,1}^{[1]} \\
Z_{1,4}=c(X_{1,4}+t_{1,2}(2))=A_{1,2}^{[4]}-A_{1,1}^{[4]} \\
Z_{2,1}=c(X_{1,1}+t_{2,2}(1))=A_{2,2}^{[1]}-A_{2,1}^{[1]} \\
Z_{2,4}=c(X_{1,4}+t_{2,2}(2))=A_{2,2}^{[4]}-A_{2,1}^{[4]}
\end{align}
These are exactly the statistics in \eqref{Z1}-\eqref{Z4}. Hence, the reliability and privacy constraints follow exactly as in Section~\ref{motivating}. The total download cost in this case is $D=4+4=8$, which is consistent with the download cost for the general case in (\ref{Download cost}), $D = \left\lceil\frac{2* 2}{2-1}\right\rceil + \left\lceil\frac{2*2}{2-1}\right\rceil = 8$.

\subsection{An Example for Heterogeneous Number of Databases}
In this example, we consider a general case, where there are no constraints on the number of databases associated with each party or on the cardinality of the sets. In this example, we have $M=4$ parties with $N_1=2$, $N_2=3$, $N_3=5$, and $N_4=4$ associated databases. The four parties have the following data sets  and the corresponding incidence vectors,
\begin{align}
&\mbox{Party}~P_1: \quad \mathcal{P}_1 = \{1,2,3,4\}, X_1 = [X_{1,1}\:\:X_{1,2}\:\:X_{1,3}\:\:X_{1,4}\:\:X_{1,5}]^T = [1\:\:1\:\:1\:\:1\:\:0]^T\\
&\mbox{Party}~P_2: \quad \mathcal{P}_2 = \{1,2,4\}, \quad X_2 = [X_{2,1}\:\:X_{2,2}\:\:X_{2,3}\:\:X_{2,4}\:\:X_{2,5}]^T = [1\:\:1\:\:0\:\:1\:\:0]^T \\
&\mbox{Party}~P_3: \quad \mathcal{P}_3 = \{1,3,4\}, \quad X_3 = [X_{3,1}\:\:X_{3,2}\:\:X_{3,3}\:\:X_{3,4}\:\:X_{3,5}]^T = [1\:\:0\:\:1\:\:1\:\:0]^T \\
&\mbox{Party}~P_4: \quad \mathcal{P}_4 = \{1,4,5\}, \quad X_4 = [X_{4,1}\:\:X_{4,2}\:\:X_{4,3}\:\:X_{4,4}\:\:X_{4,5}]^T = [1\:\:0\:\:0\:\:1\:\:1]^T 
\end{align}

First, we choose party $P_4$ for the role of the leader party, as it results in the minimum download cost $D_t=\sum_{i \neq t}\left\lceil\frac{|\cp_t|N_i}{N_i-1}\right\rceil$. Since $M=4$, we choose a retrieval field $\mathbb{F}_L$, such that $L=5$, as $L$ is the smallest prime number that satisfies $L \geq M$. 

Now, $\kappa=\max_i \: \left\lceil\frac{|\cp_4|}{N_i-1}\right\rceil=3$. Hence, for the queries, the leader $P_4$ generates $\kappa=3$ random vectors. From which, it submits $\eta_i=\left\lceil\frac{|\cp_4|}{N_i-1}\right\rceil$ to the first database associated with the $i$th party, $i=1,2,3$. Each random vector can be reused for retrieving $N_i-1$ elements from the remaining databases by adding 1 to the query vector in the positions of the desired symbols. 

Specifically, party $P_1$ has only two databases and $P_4$ is supposed to submits $\eta_1=\left\lceil\frac{3}{2-1}\right\rceil=3$ random vectors to database 1, denoted by $\mathbf{h}_\ell=[h_\ell(1) \:\: h_\ell(2) \:\: \cdots \:\: h_\ell(5)]^T$, where $\ell=1,2,3$. The leader's set is divided as $\cp_4^{11}=\{1\}$, $\cp_4^{12}=\{4\}$, and $\cp_4^{13}=\{5\}$ with $|\cp_M^{\ell_1}|=N_1-1=1$. These random vectors are generated uniformly from $\mathbb{F}_5^5$  Thus, the queries sent from $P_4$ to $P_1$ are generated as follows,
\begin{align}
Q_{1,1}^{[1]} &= [h_1(1)\quad h_1(2)\quad h_1(3)\quad h_1(4)\quad h_1(5)]^T \\
Q_{1,2}^{[1]} &= [h_1(1)+1\quad h_1(2)\quad h_1(3)\quad h_1(4)\quad h_1(5)]^T \\
Q_{1,1}^{[4]} &= [h_2(1)\quad h_2(2)\quad h_2(3)\quad h_2(4)\quad h_2(5)]^T  \\
Q_{1,2}^{[4]} &= [h_2(1)\quad h_2(2)\quad h_2(3)\quad h_2(4)+1\quad h_2(5)]^T \\
Q_{1,1}^{[5]} &= [h_3(1)\quad h_3(2)\quad h_3(3)\quad h_3(4)\quad h_3(5)]^T \\
Q_{1,2}^{[5]} &= [h_3(1)\quad h_3(2)\quad h_3(3)\quad h_3(4)\quad h_3(5)+1]^T
\end{align}

Party $P_2$ has three databases and $P_4$ only needs to send $\eta_2=\left\lceil\frac{4}{3-1}\right\rceil=2$ random vectors to database 1 of client $P_2$. Each random vector can be reused at databases 2, 3 to retrieve 2 desired symbols. The leader's set is divided as $\cp_4^{21}=\{1,4\}$, and $\cp_4^{22}=\{5\}$. Without loss of generality, $P_4$ uses $\mathbf{h}_1$ to obtain the information of $X_{2,1},X_{2,4}$ and $\mathbf{h}_2$ is used to obtain the information of $X_{2,5}$. Note that, in this case no query is needed to be sent to the third database to retrieve $X_{2,5}$. Thus, the queries sent from $P_4$ to $P_2$ are generated as follows,
\begin{align}
Q_{2,1}^{[1,4]} &= [h_1(1)\quad h_1(2)\quad h_1(3)\quad h_1(4)\quad h_1(5)]^T \\
Q_{2,2}^{[1,4]} &= [h_1(1)+1\quad h_1(2)\quad h_1(3)\quad h_1(4)\quad h_1(5)]^T \\
Q_{2,3}^{[1,4]} &= [h_1(1)\quad h_1(2)\quad h_1(3)\quad h_1(4)+1\quad h_1(5)]^T \\
Q_{2,1}^{[5]} &=  [h_2(1)\quad h_2(2)\quad h_2(3)\quad h_2(4)\quad h_2(5)]^T \\
Q_{2,2}^{[5]} &=  [h_2(1)\quad h_2(2)\quad h_2(3)\quad h_2(4)\quad h_2(5)+1]^T
\end{align}

Party $P_3$ has five databases and $P_4$ needs to send $\eta_3=\left\lceil\frac{4}{5-1}\right\rceil=1$ random vector to database 1 and reuse this vector to retrieve all the desired symbols from databases 2 through 4. Thus, the queries sent from $P_4$ to $P_3$ are generated as follows,
\begin{align}
Q_{3,1}^{[1,4,5]} &= [h_1(1)\quad h_1(2)\quad h_1(3)\quad h_1(4)\quad h_1(5)]^T \\
Q_{3,2}^{[1,4,5]} &= [h_1(1)+1\quad h_1(2)\quad h_1(3)\quad h_1(4)\quad h_1(5)]^T \\
Q_{3,3}^{[1,4,5]} &= [h_1(1)\quad h_1(2)\quad h_1(3)\quad h_1(4)+1\quad h_1(5)]^T \\
Q_{3,4}^{[1,4,5]} &=  [h_2(1)\quad h_2(2)\quad h_2(3)\quad h_2(4)\quad h_2(5)+1]^T
\end{align} 

\sloppy The clients share the following common randomness. A global randomness $c \sim \text{uniform}\{1,2,3,4\}$ is shared among all databases at all clients. A local randomness $\mathbf{s}_1=[s_1(1) \:\: s_1(2) \:\: s_1(3)]$ is shared among the databases of $P_1$, and similarly $\mathbf{s}_2=[s_2(1) \:\: s_1(2)]$, $\mathbf{s}_3=[s_3(1)]$ are shared among the databases of $P_2$ and $P_3$, respectively. The random variable $s_i(\ell) \sim \text{uniform}\{0,1,2,3,4\}$. Finally, database 2 which is associated with $P_1$, generates the individual randomness $t_{1,2}=[t_{1,2}(1) \:\: t_{1,2}(2) \:\: t_{1,2}(3)]$. Similarly, at $P_2$, database 2 generates $t_{2,2}=[t_{2,2}(1) \:\: t_{2,2}(2)]$, and database 3 generates $t_{2,3}$. Each element of the common randomness $\mathbf{t}_{i,j}$ for $i=1,2$ and $j=2,3$ is generated uniformly and independently from $\mathbb{F}_5$. The variables $(\mathbf{t}_{i,j}, \: i=1,2, \: j=2,3)$ are sent to $P_3$. The individual correlated randomness $t_{3,j}$ at $P_3$ is calculated as,
\begin{align}
t_{3,2}=2-t_{1,2}(1)-t_{2,2}(1) & \Longleftrightarrow t_{1,2}(1)+t_{2,2}(1)+t_{3,2}=2 \label{t1}\\
t_{3,3}=2-t_{1,2}(2)-t_{2,3} & \Longleftrightarrow 
t_{1,2}(2)+t_{2,3}+t_{3,3}=2  \\
t_{3,4}=2-t_{1,2}(3)-t_{2,2}(2) & \Longleftrightarrow t_{1,2}(3)+t_{2,2}(2)+t_{3,4}=2 \label{t2}
\end{align}

According to this construction, the leader receives the following answer strings from $P_1$,
\begin{align}
A_{1,1}^{[1]} &= c\left(\sum\limits_{k=1}^5 h_1(k)X_{1,k} + s_1(1)\right) \\
A_{1,2}^{[1]} &= c\left(\sum\limits_{k=1}^5 h_1(k)X_{1,k} + s_1(1) + X_{1,1} + t_{1,2}(1)\right) \\
A_{1,1}^{[4]} &= c\left(\sum\limits_{k=1}^5 h_2(k)X_{1,k} + s_1(2) \right) \\
A_{1,2}^{[4]} &= c\left(\sum\limits_{k=1}^5 h_2(k)X_{1,k} + s_1(2) + X_{1,4} + t_{1,2}(2)\right) \\
A_{1,1}^{[5]} &= c\left(\sum\limits_{k=1}^5 h_3(k)X_{1,k} + s_1(3) \right) \\
A_{1,2}^{[5]} &= c\left(\sum\limits_{k=1}^5 h_3(k)X_{1,k} + s_1(3)+ X_{1,5} + t_{1,2}(3)\right) 
\end{align}

Similarly, $P_4$ receives the following responses from $P_2$,
\begin{align}
A_{2,1}^{[1,4]} &= c\left(\sum\limits_{k=1}^5 h_1(k)X_{2,k} + s_2(1)\right) \\
A_{2,2}^{[1,4]} &= c\left(\sum\limits_{k=1}^5 h_1(k)X_{2,k} + s_2(1) + X_{2,1} + t_{2,2}(1)\right)\\
A_{2,3}^{[1,4]} &= c\left(\sum\limits_{k=1}^5 h_1(k)X_{2,k} + s_2(1) + X_{2,4} + t_{2,3}\right) \\
A_{2,1}^{[5]} &= c\left(\sum\limits_{k=1}^5 h_2(k)X_{2,k} + s_2(2)\right) \\
A_{2,2}^{[5]} &= c\left(\sum\limits_{k=1}^5 h_2(k)X_{2,k} + s_2(2) + X_{2,5} + t_{2,2}(2)\right)
\end{align}

Finally, $P_4$ receives the following responses from $P_3$,
\begin{align}
A_{3,1}^{[1,4,5]} &= c\left(\sum\limits_{k=1}^5 h_1(k)X_{3,k} + s_3(1) \right) \\
A_{3,2}^{[1,4,5]} &= c\left(\sum\limits_{k=1}^5 h_1(k)X_{3,k} + s_3(1) + X_{3,1} + t_{3,2}\right) \\
A_{3,3}^{[1,4,5]} &= c\left(\sum\limits_{k=1}^5 h_1(k)X_{3,k} + s_3(1) + X_{3,4} + t_{3,3}\right) \\
A_{3,4}^{[1,4,5]} &= c\left(\sum\limits_{k=1}^5 h_1(k)X_{3,k} + s_3(1) + X_{3,5} + t_{3,4}\right)
\end{align}

The leader party $P_4$ proceeds with decoding by removing the random responses created at database 1 of all clients $P_1$, $P_2$ and $P_3$, i.e., it constructs $Z_{i,j}$ for $i=1,2,3$ and $j=1,2,3,4,5$ by subtracting the responses $A_{i,1}$,
\begin{align}
Z_{1,1}&=c(X_{1,1} + t_{1,2}(1))= A_{1,2}^{[1]} - A_{1,1}^{[1]} \\
Z_{1,4}&=c(X_{1,4} + t_{1,2}(2)) = A_{1,2}^{[4]} - A_{1,1}^{[4]} \\
Z_{1,5}&=c(X_{1,5} + t_{1,2}(3)) = A_{1,2}^{[5]} - A_{1,1}^{[5]} \\
Z_{2,1}&=c(X_{2,1} + t_{2,2}(1)) = A_{2,2}^{[1,4]} - A_{2,1}^{[1,4]} \\
Z_{2,4}&=c(X_{2,4} + t_{2,3}) = A_{2,3}^{[1,4]} - A_{2,1}^{[1,4]} \\
Z_{2,5}&=c(X_{2,5} + t_{2,2}(2)) = A_{2,2}^{[5]} - A_{2,1}^{[5]} \\
Z_{3,1}&=c(X_{3,1} + t_{3,2}) = A_{3,2}^{[1,4,5]} - A_{3,1}^{[1,4,5]} \\
Z_{3,4}&=c(X_{3,4} + t_{3,3}) = A_{3,3}^{[1,4,5]} - A_{3,1}^{[1,4,5]} \\
Z_{3,5}&=c(X_{3,5} + t_{3,4}) = A_{3,4}^{[1,4,5]} - A_{3,1}^{[1,4,5]} 
\end{align}

The MP-PSI determination at $P_4$ concludes by evaluating the following indicators, $E_j$, for $j=1,4,5$ as,
\begin{align}
E_1 &= \sum_{i=1}^3 Z_{i,1}=c(X_{1,1} + X_{2,1} + X_{3,1} + t_{1,2}(1) + t_{2,2}(1) + t_{3,2}) \\
E_4 &= \sum_{i=1}^3 Z_{i,4}=c(X_{1,4} + X_{2,4} + X_{3,4} + t_{1,2}(2) + t_{2,3} + t_{3,3}) \\
E_5 &= \sum_{i=1}^3 Z_{i,5}=c(X_{1,5} + X_{2,5} + X_{3,5} + t_{1,2}(3) + t_{2,2}(2) + t_{3,4})
\end{align}

By observing that the sum of the correlated randomness in $E_j$ according to \eqref{t1}-\eqref{t2} is equal $2$, we note that $E_j=0$ if and only if $\sum_{i=1}^3 X_{i,j}=3$, i.e., if $X_{1,j}=X_{2,j}=X_{3,j}=1$ simultaneously. Consequently, $E_1=E_4=0$ irrespective to $c$, while $E_5 \neq 0$ and $P_4$ can reliably calculate $\cap_{i=1,2,3,4} ~ \cp_i=\{1,4\}$. On the other hand, for $E_5$, $X_{1,5} + X_{2,5} + X_{3,5} + t_{1,4} + t_{2,4} + t_{3,4}$ is equal to 2 and then $E_5$ must be one value in the set $\{1,2,3,4\}$ depending on the value of $c$. Now, we calculate the value of the expression $X_{1,5} + X_{2,5} + X_{3,5}$ from the perspective of the leader party $P_4$. If $E_5$ is $1$, $\mathbb{P}(X_{1,5} + X_{2,5} + X_{3,5} = l) = \frac{1}{4}, \forall l = \{0,1,2,3\}$ because $c$ is uniformly distributed over $\{1,2,3,4\}$. The conclusion is exactly the same when $E_5$ is equal to $2$, $3$ or $4$. Thus, the only information that $P_4$ can obtain for the element $5$ is that client parties $P_1$ , $P_2$ and $P_3$ cannot contain it at the same time. The privacy of leader party is preserved because each element in the queries is uniformly distributed over the finite field $\mathbb{F}_5$. Hence, no information about $P_4$ is leaked from the queries. The total download cost in this case is $D=6+5+4=15$, which is consistent with the download cost for the general case in (\ref{Download cost}), $D=\left\lceil\frac{3*2}{2-1}\right\rceil + \left\lceil\frac{3*3} {3-1}\right\rceil + \left\lceil\frac{3*5}{5-1}\right\rceil = 15$.

\section{Conclusion and Future Work}
We formulated the problem of MP-PSI from an information-theoretic point of view. We investigated a specific mode of communication, namely, single round communication between the leader and clients. We proposed a novel achievable scheme for the MP-PSI problem. Our scheme hinges on a careful design and sharing of randomness between client parties prior to commencing the MP-PSI operation. Our scheme is not a straightforward extension to the 2-party PSI scheme, as applying the 2-party PSI scheme $M-1$ times leaks information beyond the intersection $\cap_{i=1}^M \cp_i$. The download cost of our scheme matches the sum of download cost of pair-wise PSI despite the stringent privacy constraint in the case of MP-PSI. We note that this work provides only an achievable scheme with no claim of optimality. A converse proof is needed to assess the efficiency of our scheme. Furthermore, several interesting directions can be pursued based on this work. First, one can investigate the MP-PSI in more general communication settings (not necessarily leader-to-clients). Second, one can study the case where the communication between the parties is done over multiple rounds (in contrast to the single round of communication in this work). Third, one can investigate the case of calculating more general set functions (not necessarily the intersection). 

\bibliographystyle{unsrt}
\bibliography{reference}

\begin{thebibliography}{10}

\bibitem{PSI_first}
M.~Freedman, K.~Nissim, and B.~Pinkas.
\newblock Efficient private matching and set intersection.
\newblock In {\em International Conference on the Theory and Applications of
  Cryptographic Techniques}, pages 1--19. Springer, 2004.

\bibitem{PSI_computational}
H.~Chen, K.~Laine, and P.~Rindal.
\newblock Fast private set intersection from homomorphic encryption.
\newblock In {\em ACM SIGSAC Conference on Computer and Communications
  Security}, pages 1243--1255. ACM, 2017.

\bibitem{PSI_efficient}
D.~Dachman-Soled, T.~Malkin, M.~Raykova, and M.~Yung.
\newblock Efficient robust private set intersection.
\newblock In {\em International Conference on Applied Cryptography and Network
  Security}, pages 125--142. Springer, 2009.

\bibitem{PSI_survey}
E.~De Cristofaro and G.~Tsudik.
\newblock Practical private set intersection protocols with linear complexity.
\newblock In {\em International Conference on Financial Cryptography and Data
  Security}, pages 143--159. Springer, 2010.

\bibitem{PSIjournal}
Z.~Wang, K.~Banawan, and S.~Ulukus.
\newblock Private set intersection: A multi-message symmetric private
  information retrieval perspective.
\newblock Available at arXiv: 1912.13501.

\bibitem{PIR_ORI}
B.~Chor, E.~Kushilevitz, O.~Goldreich, and M.~Sudan.
\newblock Private information retrieval.
\newblock {\em Journal of the ACM}, 45(6):965--981, November 1998.

\bibitem{PIR}
H.~Sun and S.~A. Jafar.
\newblock The capacity of private information retrieval.
\newblock {\em IEEE Trans. on Info. Theory}, 63(7):4075--4088, July 2017.

\bibitem{JafarColluding}
H.~Sun and S.~A. Jafar.
\newblock The capacity of robust private information retrieval with colluding
  databases.
\newblock {\em IEEE Trans. on Info. Theory}, 64(4):2361--2370, April 2018.

\bibitem{arbitraryCollusion}
R.~Tajeddine, O.~W. Gnilke, D.~Karpuk, R.~Freij-Hollanti, C.~Hollanti, and
  S.~El Rouayheb.
\newblock Private information retrieval schemes for coded data with arbitrary
  collusion patterns.
\newblock In {\em IEEE ISIT}, June 2017.

\bibitem{RobustPIR_Razane}
R.~Tajeddine and S.~El Rouayheb.
\newblock Robust private information retrieval on coded data.
\newblock In {\em IEEE ISIT}, June 2017.

\bibitem{Staircase_PIR}
R.~Bitar and S.~El Rouayheb.
\newblock Staircase-{PIR}: Universally robust private information retrieval.
\newblock In {\em IEEE ITW}, pages 1--5, November 2018.

\bibitem{SPIR}
H.~Sun and S.~A. Jafar.
\newblock The capacity of symmetric private information retrieval.
\newblock {\em IEEE Transactions on Information Theory}, 65(1):322--329,
  January 2019.

\bibitem{codedsymmetric}
Q.~{Wang} and M.~{Skoglund}.
\newblock Symmetric private information retrieval from mds coded distributed
  storage with non-colluding and colluding servers.
\newblock {\em IEEE Trans. on Info. Theory}, 65(8):5160--5175, August 2019.

\bibitem{wang2017linear}
Q.~Wang and M.~Skoglund.
\newblock Linear symmetric private information retrieval for {MDS} coded
  distributed storage with colluding servers.
\newblock In {\em IEEE ITW}, pages 71--75, November 2017.

\bibitem{SPIR_Mismatched}
Q.~Wang, H.~Sun, and M.~Skoglund.
\newblock Symmetric private information retrieval with mismatched coded
  messages and randomness.
\newblock In {\em IEEE ISIT}, pages 365--369, July 2019.

\bibitem{ChaoTian_leakage}
T.~{Guo}, R.~{Zhou}, and C.~{Tian}.
\newblock On the information leakage in private information retrieval systems.
\newblock {\em IEEE Trans. on Info. Forensics and Security}, 15:2999--3012,
  2020.

\bibitem{KarimCoded}
K.~Banawan and S.~Ulukus.
\newblock The capacity of private information retrieval from coded databases.
\newblock {\em IEEE Trans. on Info. Theory}, 64(3):1945--1956, March 2018.

\bibitem{codedcolluded}
R.~Freij-Hollanti, O.~Gnilke, C.~Hollanti, and D.~Karpuk.
\newblock Private information retrieval from coded databases with colluding
  servers.
\newblock {\em SIAM Journal on Applied Algebra and Geometry}, 1(1):647--664,
  2017.

\bibitem{codedcolludingZhang}
Y.~Zhang and G.~Ge.
\newblock A general private information retrieval scheme for {MDS} coded
  databases with colluding servers.
\newblock {\em Designs, Codes and Cryptography}, 87(11), November 2019.

\bibitem{Kumar_PIRarbCoded}
S.~Kumar, H.-Y. Lin, E.~Rosnes, and A.~G. i~Amat.
\newblock Achieving maximum distance separable private information retrieval
  capacity with linear codes.
\newblock {\em IEEE Trans. on Information Theory}, 65(7):4243--4273, July 2019.

\bibitem{codedcolludingJafar}
H.~Sun and S.~A. Jafar.
\newblock Private information retrieval from {MDS} coded data with colluding
  servers: Settling a conjecture by {F}reij-{H}ollanti et al.
\newblock {\em IEEE Trans. on Info. Theory}, 64(2):1000--1022, February 2018.

\bibitem{MM-PIR}
K.~Banawan and S.~Ulukus.
\newblock Multi-message private information retrieval: Capacity results and
  near-optimal schemes.
\newblock {\em IEEE Trans. on Info. Theory}, 64(10):6842--6862, October 2018.

\bibitem{MPIRcodedcolludingZhang}
Y.~Zhang and G.~Ge.
\newblock Multi-file private information retrieval from {MDS} coded databases
  with colluding servers.
\newblock Available at arXiv: 1705.03186.

\bibitem{BPIRjournal}
K.~Banawan and S.~Ulukus.
\newblock The capacity of private information retrieval from {B}yzantine and
  colluding databases.
\newblock {\em IEEE Trans. on Info. Theory}, 65(2):1206--1219, February 2019.

\bibitem{CodeColludeByzantinePIR}
R.~Tajeddine, O.~W. Gnilke, D.~Karpuk, R.~Freij-Hollanti, and C.~Hollanti.
\newblock Private information retrieval from coded storage systems with
  colluding, {B}yzantine, and unresponsive servers.
\newblock {\em IEEE Trans. on Info. Theory}, 65(6):3898--3906, June 2019.

\bibitem{tandon2017capacity}
R.~Tandon.
\newblock The capacity of cache aided private information retrieval.
\newblock In {\em Allerton Conference}, October 2017.

\bibitem{KimCache}
M.~Kim, H.~Yang, and J.~Lee.
\newblock Cache-aided private information retrieval.
\newblock In {\em Asilomar Conference}, October 2017.

\bibitem{wei2017fundamental}
Y.-P. Wei, K.~Banawan, and S.~Ulukus.
\newblock Fundamental limits of cache-aided private information retrieval with
  unknown and uncoded prefetching.
\newblock {\em IEEE Trans. on Info. Theory}, 65(5):3215--3232, May 2019.

\bibitem{wei2017fundamental_partial}
Y.-P. Wei, K.~Banawan, and S.~Ulukus.
\newblock Cache-aided private information retrieval with partially known
  uncoded prefetching: Fundamental limits.
\newblock {\em IEEE JSAC}, 36(6):1126--1139, June 2018.

\bibitem{PIR_cache_edge}
S.~Kumar, A.~G. i~Amat, E.~Rosnes, and L.~Senigagliesi.
\newblock Private information retrieval from a cellular network with caching at
  the edge.
\newblock {\em IEEE Trans. on Communications}, 67(7):4900--4912, July 2019.

\bibitem{kadhe2017private}
S.~{Kadhe}, B.~{Garcia}, A.~{Heidarzadeh}, S.~{El Rouayheb}, and
  A.~{Sprintson}.
\newblock Private information retrieval with side information.
\newblock {\em IEEE Trans. on Info. Theory}, 66(4):2032--2043, April 2020.

\bibitem{chen2017capacity}
Z.~{Chen}, Z.~{Wang}, and S.~A. {Jafar}.
\newblock The capacity of {T}-private information retrieval with private side
  information.
\newblock {\em IEEE Trans. on Info. Theory}, 66(8):4761--4773, 2020.

\bibitem{wei2017capacity}
Y.-P. Wei, K.~Banawan, and S.~Ulukus.
\newblock The capacity of private information retrieval with partially known
  private side information.
\newblock {\em IEEE Trans. on Info. Theory}, 65(12):8222--8231, December 2019.

\bibitem{MMPIR_PSI}
S.~P. Shariatpanahi, M.~J. Siavoshani, and M.~A. Maddah-Ali.
\newblock Multi-message private information retrieval with private side
  information.
\newblock In {\em IEEE ITW}, pages 1--5, November 2018.

\bibitem{SSMMPIR_SI1}
A.~Heidarzadeh, B.~Garcia, S.~Kadhe, S.~E. Rouayheb, and A.~Sprintson.
\newblock On the capacity of single-server multi-message private information
  retrieval with side information.
\newblock In {\em Allerton Conference}, pages 180--187, October 2018.

\bibitem{SSMMPIR_SI2}
S.~Li and M.~Gastpar.
\newblock Single-server multi-message private information retrieval with side
  information.
\newblock In {\em Allerton Conference}, pages 173--179, October 2018.

\bibitem{LiConverse}
S.~{Li} and M.~{Gastpar}.
\newblock Converse for multi-server single-message {PIR} with side information.
\newblock In {\em IEEE CISS}, March 2020.

\bibitem{StorageConstrainedPIR_Wei}
Y.-P. {Wei} and S.~{Ulukus}.
\newblock The capacity of private information retrieval with private side
  information under storage constraints.
\newblock {\em IEEE Trans. on Info. Theory}, 66(4):2023--2031, April 2020.

\bibitem{PrivateComputation}
H.~Sun and S.~A. Jafar.
\newblock The capacity of private computation.
\newblock {\em IEEE Trans. on Info. Theory}, 65(6):3880--3897, June 2019.

\bibitem{mirmohseni2017private}
M.~Mirmohseni and M.~A. Maddah-Ali.
\newblock Private function retrieval.
\newblock In {\em IWCIT}, pages 1--6, April 2018.

\bibitem{PrivateSearch}
Z.~Chen, Z.~Wang, and S.~Jafar.
\newblock The asymptotic capacity of private search.
\newblock In {\em IEEE ISIT}, June 2018.

\bibitem{abdul2017private}
M.~Abdul-Wahid, F.~Almoualem, D.~Kumar, and R.~Tandon.
\newblock Private information retrieval from storage constrained databases --
  coded caching meets {PIR}.
\newblock Available at arXiv:1711.05244.

\bibitem{StorageConstrainedPIR}
M.~A. Attia, D.~Kumar, and R.~Tandon.
\newblock The capacity of private information retrieval from uncoded storage
  constrained databases.
\newblock Available at arXiv:1805.04104v2.

\bibitem{efficient_storage_ITW2019}
K.~Banawan, B.~Arasli, and S.~Ulukus.
\newblock Improved storage for efficient private information retrieval.
\newblock In {\em IEEE ITW}, August 2019.

\bibitem{Chao_storage_cost}
C.~Tian.
\newblock On the storage cost of private information retrieval.
\newblock Available at arXiv:1910.11973.

\bibitem{PIR_decentralized}
Y.-P. Wei, B.~Arasli, K.~Banawan, and S.~Ulukus.
\newblock The capacity of private information retrieval from decentralized
  uncoded caching databases.
\newblock {\em Information}, 10, December 2019.

\bibitem{heteroPIR}
K.~Banawan, B.~Arasli, Y.-P. Wei, and S.~Ulukus.
\newblock The capacity of private information retrieval from heterogeneous
  uncoded caching databases.
\newblock {\em IEEE Trans. on Info. Theory}, 66(6):3407--3416, June 2020.

\bibitem{TamoISIT}
N.~Raviv, I.~Tamo, and E.~Yaakobi.
\newblock Private information retrieval in graph-based replication systems.
\newblock {\em IEEE Trans. on Info. Theory}, 66(6):3590--3602, June 2020.

\bibitem{Karim_nonreplicated}
K.~Banawan and S.~Ulukus.
\newblock Private information retrieval from non-replicated databases.
\newblock In {\em IEEE ISIT}, pages 1272--1276, July 2019.

\bibitem{PIR_WTC_II}
K.~Banawan and S.~Ulukus.
\newblock Private information retrieval through wiretap channel {II}: Privacy
  meets security.
\newblock {\em IEEE Trans. on Info. Theory}, 66(7):4129--4149, July 2020.

\bibitem{SecurePIR}
Q.~Wang and M.~Skoglund.
\newblock On {PIR} and symmetric {PIR} from colluding databases with
  adversaries and eavesdroppers.
\newblock {\em IEEE Trans. on Info. Theory}, 65(5):3183--3197, May 2019.

\bibitem{securePIRcapacity}
Q.~Wang, H.~Sun, and M.~Skoglund.
\newblock The capacity of private information retrieval with eavesdroppers.
\newblock {\em IEEE Trans. on Info. Theory}, 65(5):3198--3214, May 2019.

\bibitem{securestoragePIR}
H.~Yang, W.~Shin, and J.~Lee.
\newblock Private information retrieval for secure distributed storage systems.
\newblock {\em IEEE Trans. on Info. Forensics and Security}, 13(12):2953--2964,
  December 2018.

\bibitem{XSTPIR}
Z.~Jia, H.~Sun, and S.~Jafar.
\newblock Cross subspace alignment and the asymptotic capacity of ${X}$-secure
  ${T}$-private information retrieval.
\newblock {\em IEEE Trans. on Info. Theory}, 65(9):5783--5798, September 2019.

\bibitem{arbmsgPIR}
H.~Sun and S.~A. Jafar.
\newblock Optimal download cost of private information retrieval for arbitrary
  message length.
\newblock {\em IEEE Trans. on Info. Forensics and Security}, 12(12):2920--2932,
  December 2017.

\bibitem{ChaoTian_coded_minsize}
R.~Zhou, C.~Tian, H.~Sun, and T.~Liu.
\newblock Capacity-achieving private information retrieval codes from mds-coded
  databases with minimum message size.
\newblock {\em IEEE Trans. on Info. Theory}, 66(8):4904--4916, August 2020.

\bibitem{MultiroundPIR}
H.~Sun and S.~A. Jafar.
\newblock Multiround private information retrieval: Capacity and storage
  overhead.
\newblock {\em IEEE Trans. on Info. Theory}, 64(8):5743--5754, August 2018.

\bibitem{KarimAsymmetricPIR}
K.~Banawan and S.~Ulukus.
\newblock Asymmetry hurts: Private information retrieval under
  asymmetric-traffic constraints.
\newblock {\em IEEE Trans. on Info. Theory}, 65(11):7628--7645, November 2019.

\bibitem{noisyPIR}
K.~Banawan and S.~Ulukus.
\newblock Noisy private information retrieval: On separability of channel
  coding and information retrieval.
\newblock {\em IEEE Trans. on Info. Theory}, 65(12):8232--8249, December 2019.

\bibitem{PIR_lifting}
R.~G.~L. D'Oliveira and S.~El Rouayheb.
\newblock One-shot {PIR}: Refinement and lifting.
\newblock {\em IEEE Trans. on Info. Theory}, 66(4):2443--2455, April 2020.

\bibitem{PIR_networks}
R.~Tajeddine, A.~Wachter-Zeh, and C.~Hollanti.
\newblock Private information retrieval over random linear networks.
\newblock Available at arXiv:1810.08941.

\bibitem{FNP04}
M.~J. Freedman, K.~Nissim, and B.~Pinkas.
\newblock Efficient private matching and set intersection.
\newblock In {\em Advances in Cryptology - EUROCRYPT 2004}, pages 1--19.
  Springer Berlin Heidelberg, 2004.

\bibitem{HV17}
C.~Hazay and M.~Venkitasubramaniam.
\newblock Scalable multi-party private set-intersection.
\newblock In {\em Public-Key Cryptography -- PKC 2017}, pages 175--203.
  Springer Berlin Heidelberg, 2017.

\bibitem{KS05}
L.~Kissner and D.~Song.
\newblock Privacy-preserving set operations.
\newblock In {\em Advances in Cryptology -- CRYPTO 2005}, pages 241--257.
  Springer Berlin Heidelberg, 2005.

\bibitem{SPIR_ORI}
Y.~Gertner, Y.~Ishai, E.~Kushilevitz, and T.~Malkin.
\newblock Protecting data privacy in private information retrieval schemes.
\newblock In {\em Thirtieth Annual ACM Symposium on Theory of Computing}, pages
  151--160. ACM, May 1998.

\end{thebibliography}
\end{document}